\documentclass[preprint]{elsarticle}

\usepackage{mathptmx}
\usepackage{helvet}
\usepackage{courier}
\usepackage{multicol}
\usepackage[bottom]{footmisc}
\usepackage{amssymb}
\usepackage{rotating}

\usepackage{epsfig}
\usepackage{latexsym}
\usepackage{graphicx}
\usepackage{color}
\usepackage{url}
\usepackage{amsfonts}
\usepackage{pgf,pgfarrows,pgfnodes,pgfautomata,pgfheaps,pgfshade}
\usepackage{tikz}
\usetikzlibrary{arrows,automata,positioning}

\usetikzlibrary{arrows,decorations.pathmorphing,backgrounds,positioning,fit,petri}
\usepackage{etex}

\usepackage{stmaryrd}
\usepackage{amsmath}

\usepackage{bussproofs}

\newdefinition{definition}{Definition}
\newdefinition{lemma}{Lemma}
\newdefinition{theorem}{Theorem}
\newdefinition{corollary}{Corollary}
\newdefinition{proposition}{Proposition}
\newdefinition{example}{Example}
\newdefinition{remark}{Remark}
\newdefinition{fact}{Fact}
\newproof{proof}{Proof}

\newcommand{\cons}{\mathcal{C}}

\newcommand{\fine}{{\mbox{ }\nolinebreak\hfill{$\Box$}}}

\newcommand{\deriv}[1]{{\mbox{${\:\stackrel{#1}{\longrightarrow}\:}$}}}

\newcommand{\Deriv}[1]{{\mbox{${\:\stackrel{#1}{\Longrightarrow}\:}$}}}
\newcommand{\derivu}[1]{{\mbox{${\:\stackrel{#1}{\longrightarrow_1}\:}$}}}
\newcommand{\derivd}[1]{{\mbox{${\:\stackrel{#1}{\longrightarrow_2}\:}$}}}

\newcommand{\nonderiv}[1]{{\mbox{${\:\not\stackrel{#1}{\longrightarrow}\:}$}}}

\newcommand{\nderiv}[1]{\nrightarrow}

\newcommand{\eqdef}{ \doteq }

\newcommand{\bigfrac}[2]{
\renewcommand{\arraystretch}{1.5}
\begin{array}{c}#1\\
\hline
#2
\end{array}}

\newcommand{\nil}{\mbox{\bf 0}}
\newcommand{\1}{\mbox{\bf 1}}

\newcommand{\const}[1]{\mbox{{\it Const}$(#1)$}}

\renewcommand{\mid}{\;\;\big|\;\;}

\newcommand{\encodings}[2]{\ensuremath{\llbracket #2 \rrbracket_{#1}}}

\begin{document}

 \pagestyle{headings}

\begin{frontmatter}
\title{Axiomatizing NFAs Generated by Regular Grammars}

\author{Roberto Gorrieri\corref{cor1}}
\ead{roberto.gorrieri@unibo.it}

\address{Dipartimento di Informatica - Scienza e Ingegneria\\
Universit\`a di Bologna,\\
Mura A. Zamboni 7, 41027 Bologna, Italy}

\cortext[cor1]{Corresponding author}

\begin{abstract}
A subclass of nondeterministic Finite Automata generated by means of regular Grammars (GFAs, for short) is introduced.
A process algebra is proposed, whose semantics maps a term to a GFA.
We prove a representability theorem: for each  GFA $N$, there exists a process algebraic term $p$ such that its semantics is a GFA
isomorphic to $N$. Moreover, we provide a concise axiomatization of language equivalence: two GFAs $N_1$ and $N_2$
recognize the same regular language if and only if the associated terms $p_1$ and $p_2$, respectively, can be equated
by means of a set of axioms, comprising 7 axioms plus 2 conditional axioms, only.
\end{abstract}


\end{frontmatter}

%
\section{Introduction}\label{intro-sec}
%

Nondeterministic finite automata (NFA for short), originally proposed in \cite{RS59}, are a well-known model of sequential 
computation (see, e.g., \cite{HMU01} for an introductory textbook), 
specifically tailored to recognize the class of {\em regular languages}, i.e., those languages that can be
described by means of {\em regular expressions} \cite{Kleene}.

In a recent paper \cite{Gor23nfa}, we have introduced a simple process algebra, called SFM1, that truly represent the class of all NFAs, up to isomorphism.
There we also showed that language equivalence over NFAs can be axiomatized by means of a rather compact axiomatization, composed of 7 axioms
plus 3 conditional axioms. 
The main aim of this paper is to improve on this result by showing that language equivalence (for regular languages) can be axiomatized by means of a very concise axiomatization, composed of 7 axioms plus 2 conditional axioms only.

\begin{figure}[t]
\centering
    \begin{tikzpicture}[shorten >=1pt,node distance=2cm,on grid,auto]
 
    
       \node[state,label={below:$A$}]            (A)               {};
       \node[state,label={below:$B$}]            (B)              [right of=A]  {};
      \node[accepting,state,label={below:$\1$}] (q1) [right of=B] {};

      \path[->] (A) edge               node {$a$} (B)
                     edge  [loop above] node {$a$}        (A)
                 (B) edge               node {$b$} (q1)
                  edge  [loop above] node {$b$}        (B)
              ;

    \end{tikzpicture}    
\caption{An NFA recognizing the language $a^+b^+$}
\label{a+b+-fig}
\end{figure}
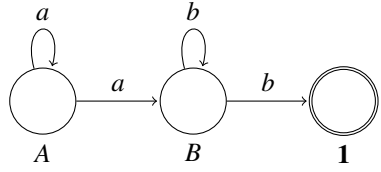

The staring point is to consider {\em regular grammars} (see, e.g., \cite{Sud97} for an introductory textbook),  i.e., those context-free grammars
whose productions have the following shape: $A \rightarrow \gamma$, where $A$ is a nonterminal symbol and
$\gamma$ can be either 
a single terminal followed by a single nonterminal (e.g., $aA$), or a single terminal (e.g., $a$), or the empty word $\epsilon$. 
For instance, the following grammar
\[ 
\begin{array}{crcl}
& A & \rightarrow & aA \mid aB
\\
& B & \rightarrow & bB \mid b 
\end{array}
\]
generating the language $L = \{a^nb^m \mid n \geq 1, m \geq 1\}$, denoted by the regular expression $a^+  b^+$,
is regular.
It is well-known \cite{Sud97,HMU01} that the class of regular grammars generates all and only the regular languages. 
Moreover, there is an obvious algorithm for mapping regular grammars over NFAs (adapted from \cite{Sud97}), described as follows. 
Given a regular grammar $G$,
the non-final states of the associated NFA $N_G$ are 
the nonterminals, the initial state of $N_G$ is the initial nonterminal symbol, 
$N_G$ may contain only one final state (we often denote by $\1$), and the transitions in $N_G$ are derived from the productions 
in $G$ as follows:
\begin{itemize}
\item transition $A \deriv{a} B$ is  in $N_G$ if $A\rightarrow aB$ is a production in $G$;
\item transition $A \deriv{a} \1$ is  in $N_G$ if $A \rightarrow a$ is a production in $G$;
\item transition $A \deriv{\epsilon} \1$ is  in $N_G$ if $A \rightarrow \epsilon$ is a production in $G$.
\end{itemize}
For instance, the NFA associated to the grammar above is outlined in Figure \ref{a+b+-fig}, following the usual drawing convention. 
It is easy to observe that the language $L(G)$ generated by the regular grammar $G$ is the same as the language $L[N_G]$ recognized by the NFA $N_G$.
Note that the NFAs produced by this algorithm have a special form, as
$(i)$ there is at most one final state, denoted by $\1$;
$(ii)$ the final state $\1$, if present, cannot be initial;
$(iii)$ there are no outgoing transitions from the final state $\1$; and, finally,
$(iv)$ all the $\epsilon$-labeled transitions have target state $\1$.

We call this kind of NFA by GFA (acronym of {\em by regular Grammar generated Finite Automaton}).
Of course, GFAs constitute a subclass of NFAs, but have the same expressive power as language acceptors: 
both classes of automata recognize all and only the regular languages.
Moreover, it is easy to observe that, given a GFA $N$, we can reconstruct a regular grammar $G_N$ as follows:
the nonterminals of $G_N$ are the non-final states of $N$, the initial nonterminal symbol is the initial state of $N$ and
\begin{itemize}
\item $q\rightarrow aq'$ is a production in $G_N$ if transition $q \deriv{a} q'$ is in $N$;
\item $q \rightarrow a$ is a production in $G_N$ if transition $q \deriv{a} \1$ is in $N$;
\item $q \rightarrow \epsilon$ is a production in $G_N$ if transition $q \deriv{\epsilon} \1$ is in $N$.
\end{itemize}
Hence, GFAs and regular grammars are in perfect correspondence, up to language equivalence.

In this paper, we propose a process algebra, called SFM$_0$, that truly represents GFAs, up to isomorphism.
More precisely, SFM$_0$, which is a slight restriction of SFM1 we proposed in  \cite{Gor23nfa} in order to axiomatize the class of NFAs, 
is a dialect of finite-state 
CCS \cite{Mil84,Mil89,GV15}, where (besides $\nil$ denoting a non-final state) an additional constant $\1$
 is introduced in order to distinguish 
the (unique) final state from the non-final ones. However, $\1$ is to be used in a constrained manner, i.e.,
only as the target of a prefix operator. 
Moreover,
SFM$_0$ comprises a disjunction operator (as for regular expressions), denoted by $\_ + \_$ (like the 
choice operator of finite-state CCS), and
a prefixing operator $\alpha.\_$ for each $\alpha \in A \cup \{\epsilon\}$, where $A$ is a 
finite alphabet and $\epsilon$ is the symbol denoting the empty word. (To be precise, the prefix $\epsilon$ can be used only with
$\1$.)
The prefixing operator is a weaker form of 
the concatenation operator $\_ \cdot \_$ of regular expressions, because it may concatenate only one single symbol
to a term, like in $a.p$. To compensate the lack of expressivity of this weaker operator, SFM$_0$ replaces the 
iteration operator $\_^*$ (or Kleene star)
of regular expressions with the more powerful recursion operator, implemented by means of process 
constants (as in CCS \cite{Mil89,GV15}), working in the same way as nonterminals are used
in regular grammars \cite{Sud97,HMU01}. For instance, consider a recursive constant $C$ defined as $C \eqdef a.C + a.\1$;
then, its semantics is a GFA with two states, the initial one (corresponding to $C$) and the
final one (corresponding to $\1$), with an $a$-labeled self-loop on the initial state and an $a$-labeled transition from the initial 
state to the final one; hence, $C$ recognizes the 
language $\{a^n \mid n \geq 1\}$, denoted by the regular expression $a^+$.

\begin{figure}[h]
\centering
\begin{tikzpicture}
  \tikzset{%
    mythick/.style={%
        line width=.35mm,>=stealth
    }
  }
  \tikzset{%
    mynode/.style={
      circle,
      fill,
      inner sep=2.1pt
    },
    shorten >= 3pt,
    shorten <= 3pt
  }
\def\eodiaglabeldist{0mm}
\def\eolabeldist{0.2mm}
\def\eofigdist{4.5cm}
\def\rrrel{$\cong_r$}
\def\eodist{0.5cm}
\def\eodisty{0.4cm}
\def\eodistw{0.8cm}

\draw [thick,fill=green!25] (-1,-1) circle [radius=1.6cm];
\draw (-1.1,-0.3) node (p0) {GFAs};
\node (p1) [mynode,below =\eodisty of p0, label={[label distance=\eodiaglabeldist]left:$N$}] {};
\node (p2) [mynode,below =\eodist of p1, label={[label distance=\eodiaglabeldist]right:$N'(p_N)$}] {};
 \path (p1) -- node (R3) [inner sep=1pt] {$\equiv$} (p2);

\draw [thick,fill=blue!25] (6,-1) circle [radius=1.6cm];
\draw (6,-0.4) node (q0) {SFM$_0$};
\node (q1) [mynode,below =\eodistw of q0, label={[label distance=\eodiaglabeldist]right:$p_N$}] {};

\draw (p1) edge[mythick,->, bend left] node[above] {compilation} (q1);
\draw (q1) edge[mythick,->, bend left] node[below] {denotational semantics} (p2);

\end{tikzpicture}
\caption{Graphical description of the representability theorem, up to isomorphism}
\label{rep-gfa}
\end{figure}
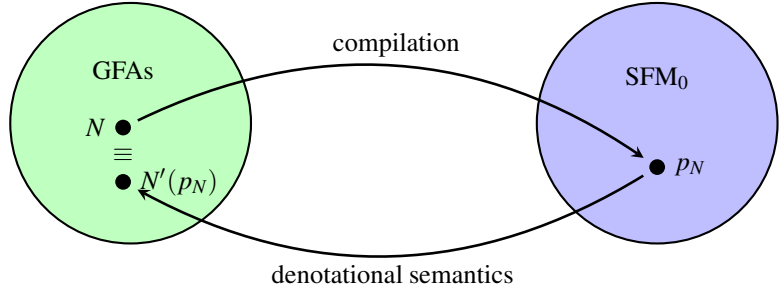

For SFM$_0$, we get
a {\em Representability Theorem} (see Figure \ref{rep-gfa}) stating that SFM$_0$ truly represents
the class of GFAs, up to isomorphism $\equiv$. In fact,
\begin{itemize}
\item[$(i)$] Given a GFA $N$, we compile it to an SFM$_0$ term representation $p_N$ by an algorithm described in the proof of Theorem \ref{representability-gfa}: so {\em all} GFAs can be represented 
by SFM$_0$; and
\item[$(ii)$] to each SFM$_0$ term $p$ (hence also to $p_N$), we associate, by means of a denotational semantics 
described in Table \ref{den-gfa-sfm-epsilon},  a GFA $N'(p)$ (hence also $N'(p_N)$): 
so {\em only} GFAs can be 
represented by SFM$_0$; and, finally,
\item[$(iii)$] the GFAs $N$ and $N'(p_N)$ are isomorphic, denoted by $N \equiv N'(p_N)$.
\end{itemize}

One advantage of this concrete representability theorem is that the algebra SFM$_0$ can be used as a basis to study
all the possible equivalences that can be defined over GFAs, not only language equivalence $\sim$ (as for regular expressions), 
but also more concrete equivalences, such as bisimulation equivalence \cite{Park81,Mil89,BBR09,GV15}, or any other 
in the 
linear-time/branching-time spectrum \cite{vG01,GV15}.

The main contribution of this paper is the proof that
language equivalence can be characterized over SFM$_0$ terms by means of a very concise axiomatization, 
comprising only 7 axioms (4 for disjunction and 3 for prefixing) and 2 conditional axioms (for recursion). 
The axioms we present are very similar to known axioms \cite{Kleene,Salomaa,Kozen,Mil84,Gor20ic}.
In fact, for disjunction they are the usual ones: the choice operator is an idempotent, commutative monoid, with $\nil$ as neutral element.
The axioms for prefixing are similar to those for concatenation: $\nil$ is an annihilator, 
the prefixing operator distributes over a choice term and $\epsilon$ can be absorbed when it prefixed by a symbol. The axioms for recursion are also similar to axioms developed for 
SFM \cite{Gor19,Gor20ic}, in turn inspired to those for finite-state CCS \cite{Mil89}: the {\em unfolding} axiom 
(stating that a constant $C$, defined as $C \eqdef p\{C/x\}$, 
can be equated to its body $p\{C/x\}$), and the {\em folding} axiom (stating that if $C \eqdef p\{C/x\}$ and 
$q = p\{q/x\}$, then $C = q$). Our axiomatization is more concise than that proposed in \cite{Gor23nfa}, because 
the prefix
$\epsilon$ is used in a very restricted manner, so that
{\em excision} axiom, stating that $\epsilon$-labeled self-loops can be removed 
(i.e., if $C \eqdef (\epsilon.x +q)\{C/x\}$ and $D \eqdef q\{D/x\}$, then $C = D$) is not necessary.

The paper is organized as follows. Section \ref{def-sec} recalls the basic definitions about GFAs, including language 
equivalence $\sim$, bisimulation equivalence $\simeq$ and isomorphism equivalence $\equiv$. 
Section \ref{sfm-epsilon-sec} introduces the algebra SFM$_0$ (i.e., its syntax and its denotational semantics 
in terms of GFAs) and proves the representability theorem sketched in Figure \ref{rep-gfa}.
Section \ref{cong-sec} shows that language equivalence $\sim$ is a congruence for all the operators of SFM$_0$, notably recursion.
Section \ref{alg-prop-sec} studies the algebraic properties of language equivalence over SFM$_0$ terms, that 
are useful to prove the soundness of the axiomatization. 
In Section \ref{axiom-sec}, we first
present the set of axioms and then prove that the axiomatization is sound and complete. 
Section \ref{conc-sec} adds concluding remarks.

%
\section{GFA: Finite Automata generated by regular Grammars} \label{def-sec}
%

Let $A$ denote a finite alphabet, ranged over by $a, b, \ldots$. Let $A^*$ denote the set of 
all the words over $A$ (including the empty word
$\epsilon$), ranged over by $\sigma$. 
Let $\alpha$ range over $A \cup \{\epsilon\}$. Let $\mathcal{P}(A^*)$ denote the set of all formal languages
over $A$, ranged over by $L$, possibly indexed.

\begin{definition} A GFA is a tuple $N = (Q, A,$ $T, F, q_0) $ where
	\begin{itemize}
		\item $Q$ is a finite set of {\em non-final} states, ranged over by $q$ (possibly indexed);
		\item $F$ is the set of {\em final} states, such that $|F| \leq 1$; 
		$Q \cap F = \emptyset$ and $Q \cup F$ is ranged over by $r$ (possibly indexed);
		\item $A$ is a finite alphabet;
		\item $T \subseteq (Q \times A \times (Q \cup F)) \; \cup \; (Q \times \{\epsilon\} \times F) $ is the set of transitions, 
		ranged over by $t$ (possibly indexed);
		\item $q_0 \in Q$ is the initial state.
	\end{itemize}
	
Given a transition $ t = (q,\alpha, r)$, $ q $ is called the \textit{source}, $ \alpha $ the \textit{label} of the 
transition (denoted also by $l(t)$), and 
$r$ the \textit{target}; this is usually represented as $q \deriv{\alpha} r$. We use the notation
$q \nonderiv{a} r$ to denote that $(q, a, r) \not \in T$.

A GFA $N = (Q, A, T, F, q_0) $ is {\em semi-deterministic} if
$\forall q\in Q \, \forall a \in A$ there exists exactly one state $q' \in Q$ such that $(q, a, q') \in T$.
A semi-deterministic GFA is called DGFA, for short.
\fine
\end{definition}

By definiton, in a GFA $N = (Q, A, T, F, q_0) $, 
$(i)$ there is at most one final state, usually denoted by $\1$ (because $|F| \leq 1$);
$(ii)$ the final state $\1$, if present, cannot be initial ($q_0 \in Q)$;
$(iii)$ there are no outgoing transitions from the final state $\1$ (because the source of a transition can only be a non-final state); and, finally,
$(iv)$ all the $\epsilon$-labeled transitions have target state $\1$ (because of the definition of $T$).

Note that a semi-deterministic GFA is {\em not deterministic}, in the usual sense; in fact, 
from a state there may be even two transitions with the same label: 
one reaching the final state and another one reaching a non-final state. 

\begin{definition}\label{def-derivstar}	\textbf{(Reachability relation and reduced GFA)}
Given $N = (Q, A, T, F, q_0)$,  the {\em reachability relation} ${\Rightarrow } \subseteq (Q \cup F) \times A^* \times (Q \cup F)$ is the least relation 
	induced by the following axiom and rules:
	
	$\begin{array}{llllllllll}
	\bigfrac{}{ r \Deriv{\epsilon} r} &\quad & \bigfrac{q \deriv{\alpha} r}{ q \Deriv{\alpha} r} & \quad & 
	 \bigfrac{q \Deriv{\sigma} q' \; \; \; q' \deriv{\alpha} r }{q \Deriv{\sigma \alpha} r} \\
	\end{array}$\\
	\noindent
	where, of course, we assume that $\epsilon$ is the identity of concatenation: 
	$\sigma \epsilon = \sigma = \epsilon \sigma$.
	We simply write $r \Rightarrow r'$ to state  that $r'$ is  {\em reachable} from $r$ when
	there exists a word $\sigma \in A^*$ such that $r \Deriv{\sigma} r'$.
	We can also define $reach(r)$ as follows: $reach(r) = \{r' \mid r \Rightarrow r'\}$.
	A GFA $N = (Q, A, T, F, q_0)$ is {\em reduced} if $reach(q_0) = Q \cup F$.
	\fine
\end{definition}

\begin{definition}\label{def-saturated-gfa}	\textbf{(Saturated GFA, $\epsilon$-free GFA)}
A GFA $N = (Q, A, T, F, q_0) $ is 
 {\em saturated} when, $\forall q \in Q$, if $q \Deriv{a} \1$, then also
$q \deriv{a} \1$; while $N$ is {\em $\epsilon$-free} when, $\forall q \in Q \setminus \{q_0\}$, we have that $q \nonderiv{\epsilon}\1$.
\fine
\end{definition}

\begin{definition}\label{def-rec-lang}	\textbf{(Recognized language and equivalence)}
Given $N = (Q, A, T, $ $F, q_0)$, the language recognized by $N$ is $L[N] = \{\sigma \in A^* \mid  q_0 \Deriv{\sigma} \1\}$.
The language recognized by state $r \in Q \cup F$ is $L[N,r] = \{\sigma \in A^* \mid r \Deriv{\sigma} \1\}$.
Two states $r_1$ and $r_2$ are  equivalent, denoted $r_1 \sim r_2$, if $L[N,r_1] = L[N,r_2]$.
Two GFAs $N_1 = (Q_1, A_1, T_1, F_1, q_{01})$ and $N_2 = (Q_2, A_2, T_2, F_2, q_{02})$ are equivalent if they recognize the same 
language, i.e., $L[N_1] = L[N_2]$.
\fine
\end{definition}

Following the informal discussion in Section \ref{intro-sec}, it is clear that there is a one-to-one correspondence between GFAs
and regular grammars \cite{Sud97}, preserving language equivalence. Therefore, the class 
of languages recognized by GFAs is the class of {\em regular languages} (see, e.g., \cite{Sud97,HMU01} for 
introductory textbooks), i.e., the languages denotable by {\em regular expressions} (originally proposed by Kleene in \cite{Kleene}; we assume the reader is familiar with this algebra).

\begin{figure}[t]
\centering
    \begin{tikzpicture}[shorten >=1pt,node distance=2cm,on grid,auto]

 
      \node[state,label={left:$q_0$}]           (q0)             {};
      \node[state,label={below:$q_1$}]           (q1)         [right of=q0]    {};
      \node[accepting,state,label={below:$r_0$}] (q2) [below of=q0] {};
    
      \path[->] (q0) edge              node {$b$} (q1)
     		 edge              node {$\epsilon$} (q2)
                     edge [loop above] node {$a$} (q0)
                     (q1) edge node {$\epsilon$} (q2)
                       edge [loop above] node {$b$} (q1)
                   ;
                 

     \node[state,label={below:$q_2$}]           (q2)     [right of=q1]      {};
      \node[accepting,state,label={below:$r_1$}] (q8) [below right of=q2] {};
      \node[state,label={below:$q_3$}]           (q3) [above right of=q8] {};
     \node[state,label={below:$q_4$}]           (q4) [right of=q3] {};
    
      \path[->] (q2) edge             node {$\epsilon$} (q8)
                           edge              node {$b$} (q3)
                    edge [loop above] node {$a$} (q2)
                    (q3) edge              node {$\epsilon$} (q8)
                    edge              node {$a$} (q4)
                     edge [loop above] node {$b$} (q3)
                     (q4)       edge [loop above] node {$a,b$} (q4)
              ;

    \end{tikzpicture}    

\caption{A GFA and a DGFA both recognizing the language $a^*b^*$}
\label{fig-nfa-rep1}
\end{figure}
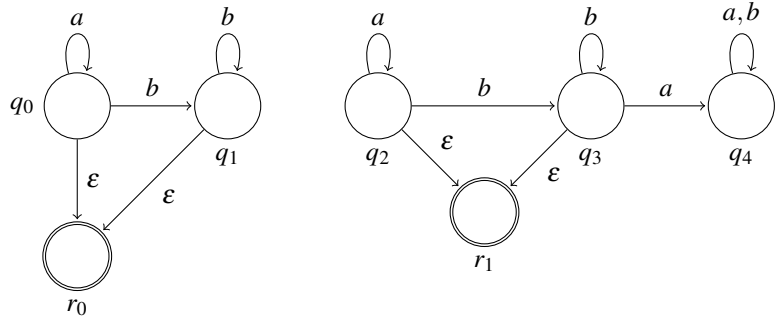

\begin{example}
Figure \ref{fig-nfa-rep1} outlines two finite automata: the left one is a GFA with initial state $q_0$ and final state $r_0$, and the right one is a DGFA 
with initial state $q_2$ and final state $r_1$. Both recognize 
the same regular language $\{a^nb^m \mid n,m \geq 0\}$, expressible by the regular expression $a^*b^*$.
As a matter of fact, $q_0 \sim q_2$ because $L[N,q_0] = L[N,q_2]$, where $N$ is the union of the two automata (union that can be done 
because the sets of states are disjoint).
\fine
\end{example}

\begin{remark}\label{satur-epsilon-equiv}
Let $N_1 = (Q_1, A_1, T_1, F_1, q_{01})$ and 
$N_2 = (Q_2, A_2, T_2, F_2, q_{02})$ be two {\em saturated} 
GFAs that are language equivalent, i.e., $L[N_1] = L[N_2]$.
If we remove all the possible $\epsilon$-labeled transitions originating from non-initial states,
we get two saturated, {\em $\epsilon$-free} GFAs $N'_1 = (Q_1, A_1, T'_1, F_1, q_{01})$ and 
$N'_2 = (Q_2, A_2, T'_2, F_2, q_{02})$ such that $L[N_1] = L[N'_1] = L[N'_2] = L[N_2]$.
In other words, if the GFAs of interest are saturated, all the $\epsilon$-labeled transitions (except 
those originating from the initial states) can be safely removed. 
\fine
\end{remark}

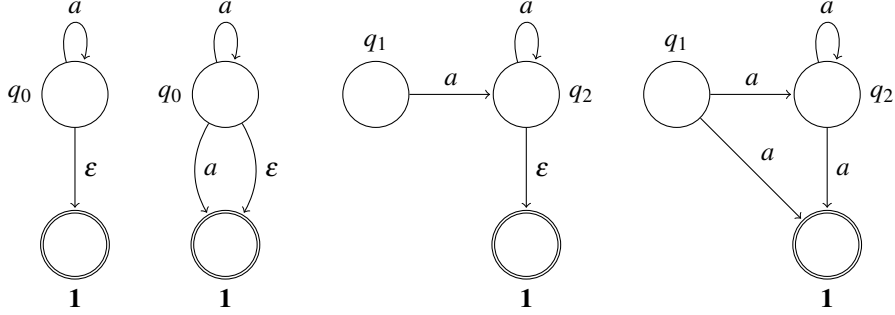
\begin{figure}[t]
\centering
    \begin{tikzpicture}[shorten >=1pt,node distance=2cm,on grid,auto]

 
      \node[state,label={left:$q_0$}]           (q0)             {};
      \node[accepting,state,label={below:$\1$}]           (q1)         [below of=q0]    {};
     
      \path[->] (q0) edge              node {$\epsilon$} (q1)
                     edge [loop above] node {$a$} (q0)
                    ;
                 

     \node[state,label={left:$q_0$}]           (q2)     [right of=q0]      {};
      \node[accepting,state,label={below:$\1$}] (q3) [below of=q2] {};
     
      \path[->] (q2) edge  [bend left]            node {$\epsilon$} (q3)
                           edge  [bend right]            node {$a$} (q3)
                    edge [loop above] node {$a$} (q2)
               ;


     \node[state,label={above:$q_1$}]           (q4)     [right of=q2]      {};
     \node[state,label={right:$q_2$}]           (q5)     [right of=q4]      {};
     \node[accepting,state,label={below:$\1$}] (q6) [below of=q5] {};
     
      \path[->] (q5) edge       node {$\epsilon$} (q6)
                    edge [loop above] node {$a$} (q5)
                    (q4) edge node {$a$} (q5)
               ;


     \node[state,label={above:$q_1$}]           (q7)     [right of=q5]      {};
     \node[state,label={right:$q_2$}]           (q8)     [right of=q7]      {};
     \node[accepting,state,label={below:$\1$}] (q9) [below of=q8] {};
     
      \path[->] (q8) edge       node {$a$} (q9)
                    edge [loop above] node {$a$} (q8)
                    (q7) edge node {$a$} (q8)
                    edge node {$a$} (q9)
               ;

    \end{tikzpicture}    

\caption{Saturation and $\epsilon$-free DGFA}
\label{fig-nfa-satur}
\end{figure}

\begin{example}
Consider the DGFA on the left of Figure \ref{fig-nfa-satur}. If we add transition $q_0 \deriv{a} \1$,
we get the {\em saturated} DGFA on its right. Note that, by saturating a DGFA $N$, we get a richer DGFA $N'$
that is language equivalent to $N$. In this case, both DGFAs recognize the language denoted by 
the regular expression $a^*$.

Now consider the DGFA in the middle of Figure \ref{fig-nfa-satur}, which recognizes the language denoted by the regular expression $a^+$. If we add the saturation transitions $q_1 \deriv{a} \1$
and $q_2 \deriv{a} \1$ and, moreover, we remove the epsilon-transition $q_2 \deriv{\epsilon} \1$, then we get 
a new DGFA, depicted on the right of Figure \ref{fig-nfa-satur}, which is $\epsilon$-free and 
language equivalent to the original one.
\fine
\end{example}

More concrete equivalences can be defined on GFAs, e.g., bisimulation equivalence \cite{Park81,Mil89,BBR09,GV15} and isomorphism equivalence.

\begin{definition}\label{def-bisf}{\bf (Bisimulation)}
Let $N = (Q, A, T, F, q_0)$ be a GFA. A {\em bisimulation} is a relation
$R\subseteq (Q\times Q) \cup (F \times F)$ such that if $(r_1, r_2) \in R$, then for all $\alpha \in A\cup\{\epsilon\} $
	\begin{itemize}
		\item $ \forall r_1' $ such that $ r_1 \deriv{\alpha} r_1' $, $ \exists r_2' $ such 
		that $ r_2 \deriv{\alpha} r_2' $ and $ (r_1', r_2') \in R $,
		\item $ \forall r_2' $ such that $ r_2 \deriv{\alpha} r_2' $, $ \exists r_1' $ such that 
		$ r_1 \deriv{\alpha} r_1' $ and $ (r_1', r_2') \in R $.
       \end{itemize}
Two states $r$ and $r'$ are bisimilar, denoted $r \simeq r'$, if there exists a bisimulation $R$ such that $(r, r') \in R$.
\fine
\end{definition}

\begin{example}
Given an NFA $N = (Q, A, T, F, q_0)$, it is easy to see that if $r_1 \simeq r_2$ then $r_1 \sim r_2$. 
The converse implication does not hold. For instance, consider Figure \ref{fig-nfa-rep1}: of 
course, $q_1 \sim q_3$ (both recognize the language $b^*$), but $q_1 \not\simeq q_3$ (as only $q_3$ has an outgoing 
$a$-labeled transition).
\fine
\end{example}

\begin{remark}\label{lang=bis-rem}
It is interesting to observe that if $N = (Q, A, T, F, q_0)$ is a saturated, $\epsilon$-free DGFA, then 
language equivalence coincides with bisimulation equivalence.
As a matter of fact, it is easy to see that for each saturated, $\epsilon$-free DGFA
the relation $R = \{(r_1, r_2) \mid r_1 \sim r_2\}$ is a bisimulation.
\fine
\end{remark}

\begin{definition}\label{def-iso-lts}\index{Isomorphism}\textbf{(Isomorphism)}
Two GFAs $N_1 = (Q_1, A_1, T_1, F_1, q_{01})$ and $N_2 = (Q_2, A_2, T_2,$ $ F_2, q_{02})$ are {\em isomorphic}, denoted by
$N_1 \equiv N_2$,
if there exists a bijection $f: Q_1 \cup F_1\rightarrow Q_2 \cup F_2$ such that:
\begin{itemize}
\item it is type-preserving: $r \in F_1$ iff $f(r) \in F_2$;
\item it preserves transitions: 
$
q\deriv{\alpha} r \in T_1 \;  \mbox{  iff  } \;  f(q)\deriv{\alpha} f(r) \in T_2
$ 
for all $q, r \in Q_1 \cup F_1$ and for all $\alpha \in A_1 \cup A_2 \cup \{\epsilon\}$; and
\item it preserves the initial states: $f(q_{01}) = q_{02}$.\\[-.9cm]
\end{itemize}
\fine
\end{definition}

Note that if $N_1 = (Q_1, A_1, T_1, F_1, q_{01})$ and $N_2 = (Q_2, A_2, T_2,$ $ F_2, q_{02})$ 
are isomorphic via $f$, then the relation $R = \{(r, f(r)) \mid r \in Q_1 \cup F_1\}$ is a bisimulation
over the union of $N_1$ and $N_2$.
Hence, isomorphism equivalence $\equiv$ is finer than bisimulation equivalence $\simeq$, in turn, 
finer than language equivalence $\sim$.

%
\section{SFM$_0$: Syntax and Semantics}\label{sfm-epsilon-sec}
%

%
\subsection{Syntax}\label{sfm-epsilon-syn-sec}
%

Let $\cons$ be a finite set of constants, disjoint from the alphabet
$A$, ranged over by $C, D, \ldots$ (possibly indexed). The size of the finite sets $A$ and $\cons$ is not important: 
we assume that they can be chosen as large as needed.  
The SFM$_0$ {\em terms} 
are generated from alphabet symbols and constants by the 
following abstract syntax:\\

$\begin{array}{llllllllllllll}
s &  ::= & \nil & | & \alpha.\1 & | & a.p & | &  s+s &  \hspace{1 cm} \mbox{{\em guarded processes}}\\
p & ::= & s & | & C & &&&&\hspace{1 cm} \mbox{{\em  processes}}\\
\end{array}$\\

\noindent
where the constant $\nil$ denotes a {\em non-final} state,
$\alpha.\1$ is a process (and a non-final state) where
the symbol $\alpha \in A \cup \{\epsilon\}$ prefixes the {\em final} state $\1$, 
$a.p$ is a process (and a non-final state) where $a \in A$ prefixes any process $p$ 
($a.-$ is a family of {\em prefixing} operators, one for each $a \in A$),
$s_1 + s_2$ denotes the disjunction of $s_1$ and $s_2$ ($- + -$ is
the {\em choice} operator), and $C$ is a constant. 
A constant $C$ may be equipped with a definition, but this must be a guarded process, i.e., 
$C \eqdef s$.

By \const{p} we denote the set of process constants $\delta(p, \emptyset)$ used by $p$, where the
auxiliary function $\delta$, which has, as an additional parameter, a set $I$ of already known constants, is 
defined as follows:\\

 $\begin{array}{rcllrclrcl}
\delta(\nil, I) & = & \emptyset & \quad &
\delta(\alpha.\1, I) & = & \emptyset \\
\delta(a.p, I) & = & \delta(p, I) & \quad &
\delta(p_1 + p_2, I) & = & \delta(p_1, I) \cup \delta(p_2, I)\\ 
\end{array}$\\
 $\begin{array}{rclrclrcl}
\quad \delta(C, I) & = & \begin{cases}
  \emptyset & \! \! \mbox{$C \in I $,}\\  
   \{C\} & \! \! \mbox{$C \not\in I  \wedge C$ undefined,} \\
  \{C\} \cup  \delta(p, I \cup \{C\})  & \! \! \mbox{$C \not\in I  \wedge C \eqdef p$.} \\
   \end{cases}
\end{array}$\\

A term $p$ is a SFM$_0$ {\em process} if  \const{p} is finite and each constant in \const{p} is equipped 
with a defining equation (in category $s$).
The set of SFM$_0$ processes is denoted by $\mathcal{P}_{SFM_0}$, while the set of its guarded processes, i.e.,
those in syntactic category $s$, by $\mathcal{P}_{SFM_0}^{grd}$.

%
\subsection{Denotational Semantics}\label{sfm-epsilon-den-sec}
%

Now we provide a construction of the GFA $\encodings{\emptyset}{p}$
associated with process $p$, which is compositional and denotational in style. The details of
the construction are outlined in Table \ref{den-gfa-sfm-epsilon}. The encoding is parametrized by a set of constants
that has  already been found while scanning $p$; such a set is initially empty and it is used to avoid 
looping on recursive constants. The definition is syntax driven
and also the states of the constructed GFA are syntactic objects, i.e., SFM$_0$ process terms (plus $\1$, which, strictly speaking, is not
an SFM$_0$ process term). 
A bit of care is needed in the rule for the choice operator: in order to include only strictly necessary 
states and transitions, the initial state $p_1$ (or $p_2$) of the GFA $\encodings{I}{p_1}$ (or $\encodings{I}{p_2}$) 
is to be kept in the GFA for $p_1 + p_2$ only if there exists a transition reaching the state $p_1$ (or $p_2$) in $\encodings{I}{p_1}$
(or $\encodings{I}{p_2}$), 
otherwise $p_1$ (or $p_2$) can be safely removed in the new GFA.
Similarly, for the rule for constants. In Table \ref{den-gfa-sfm-epsilon} we denote 
by $T(q) = \{t \in T \mid \exists \alpha \in A \cup \{\epsilon\}, \exists r\in Q \cup F. t = (q, \alpha, r)\}$  
the set of transitions in $T$ with $q$ as source state.

\begin{table}[t]
	{\renewcommand{\arraystretch}{1}
		\hrulefill\\[-1.1cm]
		
		\begin{center}\fontsize{8pt}{0.1pt}
			$\begin{array}{rcllcllcl}
			\encodings{I}{\nil}  & =  & (\{\nil\}, \emptyset, \emptyset, \emptyset, \nil) &
			\encodings{I}{\alpha.\1} \; = \;   (\{\alpha.\1\}, A, \{(\alpha.\1, \alpha, \1)\}, \{\1\}, \alpha.\1) &\\  
			&&& \mbox{where } 
			A = \begin{cases}
			\{\alpha\}  \hspace {1em} \mbox{ if $\alpha \neq \epsilon$}\\
			\emptyset \hspace{4em} \mbox{o.w.} \\
			\end{cases}\\	
			\encodings{I}{a.p}  & =  & (Q, A, T, F, a.p) & \mbox{given } 
			\encodings{I}{p}   =   (Q', A', T', F', p) \; \mbox{ and where } Q = \{a.p\} \cup Q' \\ 
			& & & F = F', \,T = \{(a.p, a, p)\} \cup T', \,
			A = \{a\} \cup A' \\	

			\encodings{I}{p_1 + p_2}  & =  & (Q, A, T, F, p_1 + p_2)& \mbox{given }  \encodings{I}{p_i}   =   
			(Q_i, A_i, T_i, F_i, p_i) \; \mbox{ for $i = 1, 2$, and where} \\  
			& & &  
			F = F_1 \cup F_2, \; Q = \{p_1 + p_2\} \cup Q_1' \cup Q_2', \mbox{ with, for $i = 1, 2$, } \\
			& & & Q'_i = \begin{cases}
			Q_i \hspace {1em} \mbox{ 
				$\exists t \in T_i$ such that $t = (q, a, p_i)$}\\  
			Q_i \setminus \{p_i\} \hspace{4em} \mbox{o.w.} \\
			\end{cases}\\
			& & &  A = A_1 \cup A_2, \;  T = T' \cup T'_1 \cup T'_2,  \mbox{ with, for $i = 1, 2$, }\\
			& & & T'_i = \begin{cases}
			T_i \hspace {4em} \mbox{
				$\exists t \in T_i. \, t = (q, a, p_i)$}\\  
			T_i \setminus T_i(p_i)  \hspace{2em} \mbox{o.w.} \\
			\end{cases}\\ & & & 
			T' = \{(p_1 + p_2, \alpha, r) \mid (p_i, \alpha, r) \in T_i, i = 1, 2\}\\
			
			\encodings{I}{C}  & =  & (\{C\}, \emptyset, \emptyset, \emptyset, C) & \mbox{if $C \in I$ } \\
			\encodings{I}{C}  & =  & (Q, A, T, F, C) & \mbox{if $C \not \in I$, given $C \eqdef p$ and } 
			\encodings{I\cup \{C\}}{p}   =   (Q', A', T', F', p) \\\ 
			&&&  A = A', F = F', Q =  \{C\} \cup Q'', \mbox{ where } \\\
			& & & Q'' =  \begin{cases}
			Q'  \hspace {1em} \mbox{
				$\exists t \in T'$ such that $t = (q, a, p)$}\\  
			Q' \setminus \{p\} \hspace{4em} \mbox{o.w.} \\ 
			\end{cases}\\
			& & & T = \{(C, \alpha, q) \mid (p, \alpha, q) \in T'\} \cup T'' \mbox{ where }\\ & & &
			T'' =   \begin{cases}
			T'  \hspace {4em} \mbox{
				$\exists t \in T'$. $t = (q, a, p)$}\\  
			T' \setminus T'(p) \hspace{4em} \mbox{o.w.} \\ 
			\end{cases}\\ 		
			\end{array}$
				
			\hrulefill
	\end{center}}
	\caption{Denotational semantics}\label{den-gfa-sfm-epsilon}
\end{table}

\begin{example}\label{ex-den-sfm1}
Consider constant $C \eqdef (a.C + \epsilon.\1) + b.D$, where $D \eqdef b.D + \epsilon.\1$. By using the definitions
in Table \ref{den-gfa-sfm-epsilon}, $\encodings{\{C,D\}}{D} = (\{D\}, \emptyset, \emptyset, \emptyset, D)$. Then, by prefixing,

$\encodings{\{C,D\}}{b.D}$ $ = (\{b.D, D\}, \{b\}, \{(b.D, b, D)\}, \emptyset, b.D)$.

Now, $\encodings{\{C,D\}}{\epsilon.\1}$ $ = (\{\epsilon.\1\}, \emptyset, \{(\epsilon.\1, \epsilon, \1)\}, \{\1\}, \epsilon.\1)$. 
Then, by summation

$\encodings{\{C,D\}}{b.D + \epsilon.\1}$ $ = (\{b.D + \epsilon.\1, D\}, \{b\}, 
\{(b.D + \epsilon.1, b, D), (b.D + \epsilon.\1, \epsilon, \1)\}, \{\1\},\\
 b.D + \epsilon.\1)$. 
Note that the states $b.D$ and $\epsilon.\1$ have been removed, as no transition in $\encodings{\{C,D\}}{b.D}$ reaches $b.D$ and
no transition in $\encodings{\{C,D\}}{\epsilon.\1}$ 
reaches $\epsilon.\1$.

\noindent
Now, the rule for constants ensures that 

$\encodings{\{C\}}{D} = (\{D\}, \{b\}, \{(D, b, D), (D, \epsilon, \1)\}, \{\1\}, D)$.

Moreover, 
$\encodings{\{C\}}{b.D} = (\{b.D, D\}, \{b\}, \{(b.D, b, D), (D, b, D), (D, \epsilon, \1)\}, \{\1\}, b.D)$.

The rule for constants also ensures that $\encodings{\{C\}}{C} = (\{C\}, \emptyset, \emptyset, \emptyset, C)$. Then, by prefixing,
$\encodings{\{C\}}{a.C} = (\{a.C, C\}, \{a\}, \{(a.C, a, C)\}, \emptyset, a.C)$, 
By summation, $\encodings{\{C\}}{a.C + \epsilon.\1} = \\
(\{a.C + \epsilon.\1, C\}, \{a\}, \{(a.C + \epsilon.\1, a, C), (a.C + \epsilon.\1, \epsilon, \1)\}, \{\1\}, a.C + \epsilon.\1)$.

By summation, again, we get
$\encodings{\{C\}}{(a.C + \epsilon.\1) + b.D} = 
(\{(a.C + \epsilon.\1) + b.D, C, D\},\\ \{a,b\},
 \{((a.C + \epsilon.\1)+b.D, a, C), ((a.C + \epsilon.\1)+b.D, \epsilon, \1),
((a.C + \epsilon.\1)+b.D, b, D),\\ (D, b, D),
 (D, \epsilon, \1)\}, \{\1\}, (a.C + \epsilon.\1)+b.D)$.

Finally, 
$\encodings{\emptyset}{C} = (\{C, D\}, \{a, b\}, \{(C, a, C), (C, \epsilon, \1), (C, b, D), (D, b, D), (D, \epsilon, \1)\}, \{\1\}, $ $ C)$. 
The resulting GFA is isomorphic to that on the left in Figure \ref{fig-nfa-rep1}, where $q_0$ corresponds to $C$, $q_1$ to $D$ and $r_0$ to $\1$.
\fine
\end{example}

\begin{theorem}\label{finite-reduced-gfa}
For each SFM$_0$ process $p$, $\encodings{\emptyset}{p}$ is a reduced GFA.	
	\proof
	By induction on the definition of $\encodings{I}{p}$. Then, the thesis follows for $I = \emptyset$.
	
	The first base case is for $\nil$ and the thesis is obvious, as,
	for each $I \subseteq \cons$, $ \encodings{I}{\nil} = (\{\nil\}, \emptyset, \emptyset, \emptyset, \nil)$, which is clearly a reduced GFA.
	Similarly for $\epsilon.\1$; in fact, for each $I \subseteq \cons$,
	$\encodings{I}{\alpha.\1} \; = \;   (\{\alpha.\1\}, A, \{(\alpha.\1, \alpha, \1)\}, \{\1\}, \alpha.\1)$
	(where $A$ can be $\emptyset$ if $\alpha = \epsilon$, or the singleton $\{\alpha\}$ otherwise) is a reduced GFA.
	The third base case is for $C$ when $C \in I$, which corresponds to the case when $C$ is not defined.
	In such a case, for each $I \subseteq \cons$ with $C \in I$, $\encodings{I}{C} =  (\{C\}, \emptyset, \emptyset, \emptyset, C)$, which
	is a reduced GFA.	
	
	The inductive cases are as follows.

	{\em Prefixing}: By induction, we can assume that $\encodings{I}{p} = (Q', A', T', F', p)$ is a reduced GFA.
		Hence, $\encodings{I}{a.p} = (Q' \cup \{a.p\}, A \cup \{a\}, T' \cup \{(a.p, a, p)\}, F', a.p)$
		is a reduced GFA as well.
		
	{\em Choice}: By induction, we can assume that $\encodings{I}{p_i} =  (Q_i, A_i, T_i, F_i, p_i)$ for $i = 1, 2$, are reduced GFAs.
	Hence, $\encodings{I}{p_1+p_2} = (Q, A, T, F, p_1 + p_2)$ is a finite automaton as well, because, according to 
	the definition in Table \ref{den-gfa-sfm-epsilon}, 
	 $|Q| \leq 1 + |Q_1| + |Q_2|$, $|A| \leq  |A_1| +  |A_2|$ and $|T| \leq |T'| + |T_1| + |T_2|$, 
	 where $|T'| \leq |T_1| + |T_2|$. Moreover, it is actually a GFA because, for $i = 1, 2,$ the set $F_i$ can 
	 only be either $\emptyset$ or $\{\1\}$, so that $F = F_1 \cup F_2$ also can be only either $\emptyset$ or
	 $\{\1\}$. Finally, the GFA $\encodings{I}{p_1+p_2}$ is reduced by construction.

	{\em Constant}: In this case, we assume that $C \eqdef p$ and that $C \not \in I$.
	By induction, we can assume that $\encodings{I\cup \{C\}}{p} =  (Q', A', T', F', p)$ is a reduced GFA.
	Hence, $\encodings{I}{C} = (Q, A, T, F, C)$ is a GFA as well, because, according to the definition 
	in Table \ref{den-gfa-sfm-epsilon}, $|Q| \leq 1 + |Q'|$, $A = A'$, $F = F'$ and $|T| \leq 2 \cdot |T'|$; and also reduced 
	by construction.
	\fine
\end{theorem}

\begin{theorem}\label{representability-gfa}
{\bf (Representability)} For each reduced GFA $N$, there exists an SFM$_0$ process $p$ such that $\encodings{\emptyset}{p}$
is isomorphic to $N$. 
\proof
Let $N = (Q, A, T, F, q_0) $ be a reduced GFA, with $Q = \{q_0, q_1, \ldots, q_m\}$. 
For $i = 0, \ldots, m$, let $T_n(q_i) = \{t \in T \mid \exists a \in A , \exists q_k \in Q. t = (q_i, a, q_k)\}$ be 
the set of transitions in $T$ with $q_i$ as source state and with a {\em non-final} target state. Moreover,
in case $F$ is not empty (say $F = \{r\}$), let also
$T_f(q_i) = \{t \in T \mid \exists \alpha \in A \cup \{\epsilon\}. t = (q_i, \alpha, r)\}$  be 
the set of transitions in $T$ with $q_i$ as source state and $r$ as {\em final} target state.

We define a process constant $C_i$ in correspondence
with state $q_i$, for $i = 0, 1, \ldots, m$, defined as follows:
\begin{itemize}
\item if $T_n(q_i) = \emptyset$ and $T_f(q_i) = \emptyset$ (i.e., $q_i$ is a deadlock), then $C_i \eqdef \nil$; 
\item if $T_n(q_i) \neq \emptyset$ and $T_f(q_i) = \emptyset$, then $C_i \eqdef \Sigma_{(q_i, a, q_k) \in T_n(q_i)} a.C_k$; 
\item if $T_n(q_i) = \emptyset$ and $T_f(q_i) \neq \emptyset$, then $C_i \eqdef \Sigma_{(q_i, \alpha, r) \in T_f(q_i)} \alpha.\1$; 
\item otherwise, $C_i \eqdef \Sigma_{(q_i, a, q_k) \in T_n(q_i)} a.C_k \, + \, \Sigma_{(q_i, \alpha, r) \in T_f(q_i)} \alpha.\1$.
\end{itemize}
Let us consider $\encodings{\emptyset}{C_0}$. It is not difficult to see
that $reach(C_0) = \{C_0, C_1 \ldots, C_m\} \cup \{\1\}$ (where state $\1$ is present only if  $F \neq \emptyset$)
because $N$ is reduced. So, the bijection we are looking for is
$f: Q \cup F \rightarrow \{C_0, C_1 \ldots, C_m\} \cup \{ \1 \}$, 
defined as $f(r) = \1$ and $f(q_i) = C_i$ for $i = 0, \ldots, m$.
It is also easy to observe that the three conditions of isomorphism are satisfied, namely:
\begin{itemize}
\item $r \in F\;$ iff $\; f(r)= \1$, and 
\item $q\derivu{\alpha}r \in N \;$ iff $ \; f(q)  \derivd{\alpha}  f(r) \in \encodings{\emptyset}{C_0}$, and
\item $C_{0} = f(q_{0})$. 
\end{itemize}
Hence, $f$ is indeed a GFA isomorphism.
\fine
\end{theorem}

\begin{corollary}\label{cor-rep}
SFM$_0$ represents, up to isomorphism, the class of reduced GFAs.
\proof
By Theorem \ref{finite-reduced-gfa} the semantics of an SFM$_0$ process is a reduced GFA; hence, 
only reduced GFAs can be represented by SFM$_0$ processes. By Theorem \ref{representability-gfa} each reduced GFA
 can be represented, up to isomorphism, by a suitable SFM$_0$ process; hence, all reduced GFAs can be represented 
 by SFM$_0$ processes.
\fine
\end{corollary}

\begin{figure}[t]
\centering
    \begin{tikzpicture}[shorten >=1pt,node distance=2cm,on grid,auto]

 
      \node[state,label={below:$q_5$}]           (q5)             {};
      \node[accepting,state,label={below:$r_0$}] (q6) [right of=q5] {};
    
      \path[->] (q5) edge              node {$a$} (q6)
                     edge [loop above] node {$b$} (q5);
                 

     \node[state,label={below:$q_6$}]           (q7)     [right of=q6]      {};
      \node[accepting,state,label={above:$r_1$}] (q9) [below right of=q7] {};
      \node[state,label={below:$q_7$}]           (q8) [above right of=q9] {};
    
      \path[->] (q7) edge              node {$a$} (q8)
                     edge              node {$b$} (q9)
                     edge [loop above] node {$b$} (q7);

    \end{tikzpicture}    

\caption{Two GFAs for Example \ref{rep-ex}}
\label{fig-nfa-rep2}
\end{figure}
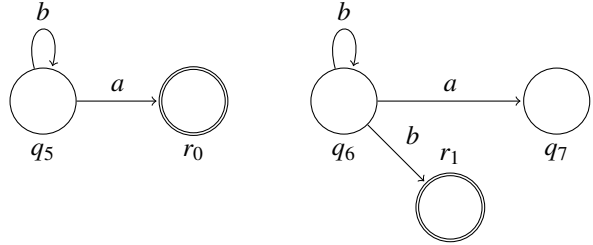

\begin{example}\label{rep-ex}
Let us consider the GFA on the left in Figure \ref{fig-nfa-rep1}. According to the construction in the proof of Theorem \ref{representability-gfa}, its SFM$_0$ representation is as follows:

$\begin{array}{rcl}
C_0 & \eqdef & a.C_0 + b.C_1 + \epsilon.\1\\
C_1 & \eqdef & b.C_1 + \epsilon.\1 \\
\end{array}$

\noindent
Let us consider the DGFA in Figure \ref{fig-nfa-rep1}. Its SFM$_0$ representation is as follows:

$\begin{array}{rcl}
C_2 & \eqdef & a.C_2 + b.C_3 + b.\1 + \epsilon.\1\\
C_3 & \eqdef & b.C_3 + b.\1 + a.C_4\\
C_4 & \eqdef & a.C_4 + b.C_4
\end{array}$

Note that if the GFA is not reduced, then we cannot find an SFM$_0$ process that represents it. For instance, if we take 
the union of the two automata in Figure \ref{fig-nfa-rep1} with initial state $q_0$, then $C_0$ represents only 
the GFA on the left.

\noindent
If we consider the first GFA in Figure \ref{fig-nfa-rep2}, we get:

$\begin{array}{rcl}
C_5 & \eqdef & a.\1 + b.C_5\\
\end{array}$

\noindent
Finally, if we consider the second GFA in Figure \ref{fig-nfa-rep2}, we get:

$\begin{array}{rcl}
C_6 & \eqdef & b.C_6 + a.C_7 + b.\1\\
C_7 & \eqdef & \nil\\
\end{array}$\\[-.4cm]
\fine
\end{example}

%
\section{Congruence}\label{cong-sec}
%

In the following, for each SFM$_0$ process $p$, its associated language $L(p)$ is the language recognized by
the reduced GFA $\encodings{\emptyset}{p}$, computed according to the denotational semantics 
in Table \ref{den-gfa-sfm-epsilon}.

\begin{definition}
Two SFM$_0$ processes $p$ and $q$ are {\em language equivalent}, denoted $p \sim q$, if
$L(p) = L(q)$, i.e., they recognize the same language.
\fine
\end{definition}

\begin{proposition}\label{prop-obvious} The following hold:

\noindent
1) $L(\nil) = \emptyset$.

\noindent
2) $L(\alpha.\1) = \{\alpha\}$.

\noindent
3) For each $p \in  \mathcal{P}_{SFM_0}$, for all $a \in A$, we have that 
$L(a.p) = \{a w \mid w \in L(p)\}$.

\noindent
4) For each $p, q \in  \mathcal{P}_{SFM_0}^{grd}$, $L(p+q) = L(p) \cup L(q)$.

\proof Obvious by construction in Table \ref{den-gfa-sfm-epsilon}.
\fine
\end{proposition}

Now we show that language equivalence is a congruence for all the SFM$_0$ operators.

\begin{proposition}\label{prop-cong1}
1) For each $p, q \in  \mathcal{P}_{SFM_0}$, if $p \sim q$, then $a.p  \sim  a.q$
for all $a \in A$.  

\noindent
2) For each $p, q \in  \mathcal{P}_{SFM_0}^{grd}$, if $p \sim q$, 
then $p + r  \sim  q + r \;$
for all $r \in  \mathcal{P}_{SFM_0}^{grd}$.

\proof By hypothesis, we know that $L(p) = L(q)$. Then the thesis follows easily by application of Proposition \ref{prop-obvious}.

Now, for case 1, $L(a.p) = \{a w \mid w \in L(p)\} = \{a w \mid w \in L(q)\} = L(a.q)$ and so
$a.p  \sim  a.q$.
While, for case 2, $L(p+r) = L(p)\cup L(r) = L(q) \cup L(r) = L(q+r)$ and so  
$p + r  \sim  q + r$.
\fine
\end{proposition}

Note that the symmetric case $r + p \sim r + q$ is implied by the fact that 
the operator of choice is commutative w.r.t. $\sim$ (see Proposition \ref{aci-+}).

Still there is one construct missing: recursion, defined over guarded terms only.
Consider an extension of SFM$_0$ where terms can be constructed using variables, such as $x, y, \ldots$
(which are in syntactic category $p$): 
this defines
an ``open'' SFM$_0$ that has the following syntax:\\

$\begin{array}{lccccccccccl}
s &  ::= & \nil & | & \alpha.\1 & | & a.p & | &  s+s & \hspace{1 cm}\\
p & ::= & s & | & C & | & x &&\hspace{1 cm}\\
\end{array}$\\

\noindent
where the semantic of $x$ is the GFA $(\{x\}, \emptyset, \emptyset, \emptyset, x)$.

Sometimes we use the notation $p(x_1, \ldots, x_n)$ to state explicitly that term $p$ is open on the tuple of variables $(x_1, \ldots, x_n)$.
For instance, $p_1(x) = a.(b.\1 + c.x)$ and $p_2(x) = a.c.x + a.b.\1$ are open guarded SFM$_0$ terms.
Similarly, $C_1(x) \eqdef a.x$ denotes explicitly that the constant $C_1$ is defined by a term open on $x$.

An open term $p(x_1, \ldots, x_n)$ can be {\em closed} by means of a substitution as follows:
\[
p(x_1, \ldots, x_n)\{r_1 / x_1, \ldots, r_n / x_n\}
\]
\noindent
with the effect that each occurrence of the free variable $x_i$ is replaced by the {\em closed} SFM$_0$
process $r_i$, for $i = 1, \ldots, n$. For instance,
$p_1(x)\{d.\nil / x\} = a.(b.\1 + c.d.\nil)$, as well as $C_1(x)\{d.\nil / x\}$ is $C_1 \eqdef a.d.\nil$ (i.e., we close the open
term $a.x$ defining the body of $C_1$).
A natural extension of language equivalence $\sim$ over open {\em guarded} terms is as follows:
$p(x_1, \ldots, x_n) \sim q(x_1, \ldots, x_n)$
if for all tuples of closed SFM$_0$ terms
$(r_1, \ldots, r_n)$, \\

$p(x_1, \ldots, x_n)\{r_1 / x_1, \ldots, r_n / x_n\}$ $\sim$ $q(x_1, \ldots, x_n)\{r_1 / x_1, \ldots, r_n / x_n\}$.\\

\noindent
E.g., it is easy to see that $p_1(x) \sim p_2(x)$. As a matter of fact, for all $r$, 
$p_1(x)\{r / x\} = a.(b.\1 + c.r)$ $\sim$ $a.c.r + a.b.\1$ = $p_2(x)\{r / x\}$, which can be easily 
proved by means of the algebraic properties (see the next section) and the congruence ones of $\sim$.
	
For simplicity's sake, let us now restrict our attention to open guarded terms using a single undefined variable.
We can {\em recursively close} an open term $p(x)$ by means of a recursively defined constant. For instance, 
$C \eqdef p(x)\{C / x\}$. The resulting process constant $C$ is a closed SFM$_0$ process. 
By saying that language equivalence is a congruence for
recursion we mean the following: If $p(x) \sim q(x)$ and $C \eqdef p(x)\{C / x\}$ and $D \eqdef q(x)\{D / x\}$, 
then $C \sim D$.
The following theorem states this fact. 
For simplicity's sake, in the following a term $p$, open on a variable $x$, is annotated 
as $p(x)$ only when it is unclear from the context, because we work with real terms and not with annotated ones.

\begin{remark}
Throughout the paper, it is assumed that whenever a constant $C$ is defined by $C \eqdef p\{C/x\}$,
then the open guarded term $p$ does not 
contain occurrences of $C$, i.e., $C \not \in \const{p}$,
so that all the instances of $C$ in $p\{C/x\}$ are due to substitution of the constant $C$ for the variable $x$. 
\fine
\end{remark}

\begin{theorem}\label{recurs-cong-th}\index{Congruence}
	Let $p$ and $q$ be two open guarded SFM$_0$ terms, with one variable $x$ at most. 
	Let $C \eqdef p\{C / x\}$, $D \eqdef q\{D / x\}$
	and $p \sim q$. Then $C \sim D$.

	\proof For the term $p$, open on $x$, we define the following languages:
	\begin{itemize}
	\item $L_p^{\downarrow}= \{\sigma \in A^* \mid p \Deriv{\sigma} \1\}$ is the set of the words recognized by $p$;
         \item $L_p^x= \{\sigma \in A^* \mid p \Deriv{\sigma} x\}$ is the set of its maximal words reaching the variable $x$.
  	\end{itemize}
It is rather easy to see that for $C \eqdef p\{C / x\}$, we have that  $L(C) = (L_p^x)^* \cdot L_p^{\downarrow}$, 
where $\_ \cdot \_$ is the concatenation operator and $\_^*$ is the Kleene star operator over languages \cite{HMU01}.
	It is not difficult to see that $p \sim q$ if and only if $L_p^{\downarrow} = L_q^{\downarrow}$ and $L_p^x = L_q^x$, 
	from which the thesis $C \sim D$ follows immediately.
		\fine
\end{theorem}

\begin{example}
Consider $p_1(x) = a.(b.\1 + c.x)$ and $p_2(x) = a.c.x + a.b.\1$. Of course, they are language equivalent and indeed
$L_{p_1}^{\downarrow} = \{ab\} = L_{p_2}^{\downarrow}$ and $L_{p_1}^x = \{ac\} = L_{p_2}^x$. 
If we consider $C \eqdef p_1\{C / x\}$ and
$D \eqdef p_2\{D / x\}$, then we get that $C \sim D$, as both recognize the regular language $(ac)^*ab$.
\fine
\end{example}

The generalization to open terms over a set $\{x_1, \ldots, x_n\}$ of variables is rather obvious. 
Consider a set of pairs of equivalent open terms of this form:\\

$\begin{array}{lrcllllll}
\mbox{$p_1(x_1, \ldots, x_n) \sim q_1(x_1, \ldots, x_n) $}\\
\mbox{$p_2(x_1, \ldots, x_n) \sim q_2(x_1, \ldots, x_n) $}\\
\mbox{$\ldots$}\\
\mbox{$p_n(x_1, \ldots, x_n) \sim q_n(x_1, \ldots, x_n) $}\\
\end{array}$\\

\noindent
where, for $i = 1, \ldots, n$, $p_i(x_1, \ldots, x_n) \sim q_i(x_1, \ldots, x_n)$ if and only if $L_{p_i}^{\downarrow} = L_{q_i}^{\downarrow}$
and $L_{p_i}^{x_j} = L_{q_i}^{x_j}$ for $j = 1, \ldots, n$.
Then, we can recursively close these terms as follows:\\

$\begin{array}{lrcllllll}
\mbox{$C_1 \eqdef p_1\{C_1/x_1, \ldots, C_n/x_n\} \hspace {3em} D_1 \eqdef q_1\{D_1/x_1, \ldots, D_n/x_n\}$}\\
\mbox{$C_2 \eqdef p_2\{C_1/x_1, \ldots, C_n/x_n\} \hspace {3em} D_2 \eqdef q_2\{D_1/x_1, \ldots, D_n/x_n\}$}\\
\mbox{$\ldots$}\\
\mbox{$C_n \eqdef p_n\{C_1/x_1, \ldots, C_n/x_n\} \hspace {3em} D_n \eqdef q_n\{D_1/x_1, \ldots, D_n/x_n\}$}\\
\end{array}$\\

\noindent
Then, the thesis is that $C_i \sim D_i$ for all $i = 1, 2, \ldots, n$.

Finally, a comment on the more concrete equivalences we have defined in Section \ref{def-sec}. Isomorphism equivalence is the most concrete
and it is not a congruence for the choice operator. For instance, $a.\nil$ generates a GFA isomorphic to the one for $a.(\nil+\nil)$; however,
if we consider the context $\_ + a.\nil$ we get two terms generating not isomorphic GFAs: $a.\nil + a.\nil$ generates a two-state GFA, 
while $a.(\nil + \nil) + a.\nil$ generates a three-state GFA.
On the contrary, it is possible to prove that bisimulation equivalence $\simeq$, which is strictly finer than language equivalence 
$\sim$, is a congruence for all the operators of SFM$_0$, by means of standard proof techniques (see, e.g., \cite{Mil89,GV15,BBR09,San10,Gor20ic}).

%
\section{Algebraic Properties}\label{alg-prop-sec}
%

Now we list some algebraic properties of language equivalence and discuss whether they hold for bisimulation equivalence or isomorphism equivalence.

\begin{proposition}\label{aci-+}{\bf (Laws of the choice op.)}
For each $p, q, r \in  \mathcal{P}_{SFM_0}^{grd}$ , the following hold:
 
 $\begin{array}{lrcllllll}
\qquad & p + (q + r)  &\sim & (p + q) + r & \quad &\; \mbox{(associativity)}\\
\qquad & p + q  &\sim & q + p & \quad &\;  \mbox{(commutativity)}\\
\qquad & p + \nil  &\sim & p &  &\;  \mbox{(identity)}\\
\qquad & p + p  &\sim & p &  & \; \mbox{(idempotency)}
\end{array}$
\proof The proof follows directly from Proposition \ref{prop-obvious}(4). For instance,
$p + q  \sim  q + p$ holds because $L(p+q) = L(p) \cup L(q) = L(q) \cup L(p) = L(q+p)$.
\fine
 \end{proposition}
 
The laws of the choice operator hold also for bisimilarity \cite{Mil89,GV15,BBR09,San10},
 but they do not hold for isomorphism equivalence. For instance, consider $p = a.A$, with $A \eqdef a.(b.B + a.A)$,
 and $q = b.B$, with $B \eqdef b.B + \epsilon.\1$; then, it is easy to see that the GFA for $p + q$ (five states) 
 is not isomorphic to the GFA for $q + p$ (four states). 
 
 \begin{proposition}\label{pref-alg-prop}{\bf (Laws of the prefixing operator)}
 For each $p, q \in  \mathcal{P}_{SFM_0}^{grd}$,
 for each $a \in A$, the following hold:
 
 $\begin{array}{lrcllllll}
\qquad & a.\nil  &\sim & \nil & \quad &\; \mbox{(annihilation)}\\
\qquad & a.(p + q)  &\sim & a.p + a.q & \quad &\;  \mbox{(distributivity)}\\
\qquad & a.\epsilon.\1  &\sim & a.\1 &  & \; \mbox{($\epsilon$-absorption)}
\end{array}$
\proof The proof follows directly from Proposition \ref{prop-obvious}. The law
$a.\nil  \sim  \nil$ holds because $L(a.\nil) = \{a w \mid w \in L(\nil)\} =
\{a w \mid w \in \emptyset\} = \emptyset = L(\nil)$. The law $a.(p + q)  \sim a.p + a.q$
holds because $L(a.(p + q)) = \{a w \mid w \in L(p+q)\} =$ $ \{a w \mid w \in L(p) \cup L(q)\} =
\{a w \mid w \in L(p)\} \cup \{a w \mid w \in L(q)\} $ $= L(a.p) \cup L(a.q) =  L(a.p + a.q)$. 
The law 
$a.\epsilon.\1  \sim a.\1$ holds because $L(a.\epsilon.\1) = \{a\} = L(a.\1)$.
\fine
 \end{proposition}

It is easy to see that none of the laws for prefixing hold for bisimilarity, and so not even for isomorphism equivalence.

We now focus on the properties of constants. First, the unfolding law is discussed, then also
the folding law, whose proof relies on some auxiliary technical machinery (fixed point theorems on a
complete lattice, see, e.g., \cite{DP02,San10}).

\begin{proposition}\label{rec-law-unf}{\bf (Unfolding)}
 For each $p \in  \mathcal{P}_{SFM_0}^{grd}$ and each $C \in \cons$, 
if $C \eqdef p$ then $C   \sim  p$.
\proof Trivial by construction in Table \ref{den-gfa-sfm-epsilon}.
\fine
 \end{proposition}

The unfolding law also holds for bisimilarity \cite{Mil89,GV15,Gor19}, but it does not hold for isomorphism equivalence.
For instance, consider $C \eqdef a.a.C + \epsilon.\1$: it is easy to see that the GFA for $C$ has three states, while the 
GFA for $a.a.C + \epsilon.\1$
has four states.

In order to prove the folding law below, 
we first observe that $(\mathcal{P}(A^*), \subseteq)$ is a {\em complete} lattice, i.e., with the property that
all the subsets of $\mathcal{P}(A^*)$ ( e.g., $\{L_i\}_{i\in I}$) have joins (i.e., $\bigcup_{i \in I} L_i$) and meets
(i.e., $\bigcap_{i \in I} L_i$), whose top element $\top$ is $A^*$ and bottom element $\bot$ is $\emptyset$.

The folding law states that if $C \eqdef p\{C/x\}$ and $q \sim p\{q/x\}$, then $ C  \sim q$.
Of course, we can easily realize that $L(C) = L_p^{\downarrow} \cup (L_p^x \cdot L(C))$ as well as
 $L(q) = L_p^{\downarrow} \cup (L_p^x \cdot L(q))$. Therefore, both $L(C)$ and $L(q)$ are a fixed point of the
 mapping $T(W) = L_p^{\downarrow} \cup (L_p^x \cdot W)$. We want to show that 
 there exists exactly one fixed point of
 $T$, so that $L(C) = L(q)$ and so 
 $C \sim q$, as required. 
 
If the mapping $T$ is {\em continuous} (i.e.,
for all increasing sequences $L_0 \subseteq L_1 \subseteq L_2 \ldots$ we have that $T(\bigcup_{i} L_i) = \bigcup_i T(L_i)$),
then the {\em least fixed point} {\tt lfp}$(T) = \bigcup_{n \geq 0} T^n(\emptyset)$, where $T^0(L) = L$ and $T^{n+1}(L) = T(T^n(L))$.
Indeed, $T$ is continuous because $T(\bigcup_{i} L_i) = L_p^{\downarrow} \cup (L_p^x \cdot \bigcup_{i} L_i)$
$= L_p^{\downarrow} \cup \bigcup_{i} (L_p^x \cdot L_i) = $ $\bigcup_{i} ( L_p^{\downarrow} \cup (L_p^x \cdot L_i)) = \bigcup_i T(L_i)$.
The least fixed point {\tt lfp}$(T) = \bigcup_{n \geq 0} T^n(\emptyset)$ can be computed as follows:\\

$\begin{array}{rcl}
T^0(\emptyset) & = & \emptyset\\
T^1(\emptyset) & = & L_p^{\downarrow}\\
T^2(\emptyset) & = & L_p^{\downarrow} \cup (L_p^x \cdot L_p^{\downarrow})\\
T^3(\emptyset) & = & L_p^{\downarrow} \cup (L_p^x \cdot L_p^{\downarrow}) \cup (L_p^x \cdot (L_p^{\downarrow} \cup (L_p^x \cdot L_p^{\downarrow})) = L_p^{\downarrow} \cup (L_p^x \cdot L_p^{\downarrow}) \cup L_p^x \cdot (L_p^x \cdot L_p^{\downarrow})\\
\ldots\\
\end{array}$

\noindent whose union is the language {\tt lfp}$(T) = (L_p^x)^* \cdot L_p^{\downarrow}$.

Dually, if $T$ is {\em co-continuous} (i.e., 
for all decreasing sequences $L_0 \supseteq L_1 \supseteq L_2 \ldots$ we have that $T(\bigcap_{i} L_i) = \bigcap_i T(L_i)$),
then the {\em greatest fixed point} {\tt gfp}$(T) = \bigcap_{n \geq 0} T^n(A^*)$.
Indeed, $T$ is co-continuous because $T(\bigcap_{i} L_i) = L_p^{\downarrow} \cup (L_p^x \cdot \bigcap_{i} L_i)$
$= L_p^{\downarrow} \cup \bigcap_{i} (L_p^x \cdot L_i) = $ $\bigcap_{i} ( L_p^{\downarrow} \cup (L_p^x \cdot L_i)) = \bigcap_i T(L_i)$.
The greatest fixed point {\tt gfp}$(T) = \bigcap_{n \geq 0} T^n(A^*)$ can be computed as follows:\\

$\begin{array}{rcl}
T^0(A^*) & = & A^*\\
T^1(A^*) & = & L_p^{\downarrow} \cup (L_p^x \cdot A^*)\\
T^2(A^*) & = & L_p^{\downarrow} \cup (L_p^x \cdot (L_p^{\downarrow} \cup (L_p^x \cdot A^*)) = L_p^{\downarrow} \cup (L_p^x \cdot L_p^{\downarrow}) \cup L_p^x \cdot (L_p^x \cdot A^*)\\
T^3(A^*) & = & L_p^{\downarrow} \cup (L_p^x \cdot 
(L_p^{\downarrow} \cup (L_p^x \cdot L_p^{\downarrow}) \cup L_p^x \cdot (L_p^x \cdot A^*)) = $ $\\
& & L_p^{\downarrow} \cup (L_p^x \cdot L_p^{\downarrow}) \cup L_p^x \cdot (L_p^x \cdot L_p^{\downarrow}) \cup 
L_p^x \cdot (L_p^x \cdot (L_p^x \cdot A^*))\\
\ldots\\
\end{array}$

\noindent whose intersection is the language {\tt gfp}$(T) = (L_p^x)^* \cdot L_p^{\downarrow}$, unless $L_p^x$ does contain
 the empty word $\epsilon$, because in such a case the application of $T$ does not restrict the newly computed language,
 so that $T^n(A^*) = A^*$ for all $n \geq 0$, and so the greatest fixed point of $T$ is the top element
 $A^*$ of the complete lattice. However, by syntactic definition, the prefix $\epsilon$ can be only used with target $\1$, so that
 $\epsilon \not\in L_p^x$.
 Therefore, by the argument above we can conclude the following.

\begin{proposition}\label{rec-law-fold}{\bf (Folding)}
 For each $p \in  \mathcal{P}_{SFM_0}^{grd}$ (open on $x$), and each $C \in \cons$, 
 if $C \eqdef p\{C/x\}$ and $q \sim p\{q/x\}$, then $ C  \sim q$.
 \proof The least fixed point and the greatest fixed point of the function
 $T(W) = L_p^{\downarrow} \cup (L_p^x \cdot W)$ do coincide, so that the fixed point is unique.
 Since both $L(C)$ and $L(q)$ are fixed points for $T$, then $L(C) = L(q)$.
 \fine
 \end{proposition}

It is possible to prove, with a standard bisimulation-based proof technique \cite{Mil84,Gor19}, that the more concrete 
bisimulation equivalence $\simeq$  satisfies the
folding law: For each $p \in  \mathcal{P}_{SFM_0}^{grd}$ (open on $x$), and each $C \in \cons$, 
 if $C \eqdef p\{C/x\}$ and $q \simeq p\{q/x\}$, then $ C  \simeq q$.

%
\section{Axiomatization}\label{axiom-sec}
%
 
 In this section we present a sound and complete, finite axiomatization of language equivalence over SFM$_0$, i.e., 
 we show that it is possible to prove syntactically, i.e., by means of an equational deductive proof, when 
 two SFM$_0$ processes recognize the same regular language. 
For a description of how equational deduction is implemented in a process algebra with process constants 
we refer the reader to \cite{Gor23nfa}.

For simplicity's sake, the syntactic definition of {\em open} (i.e., with variables) SFM$_0$ is given with only 
one syntactic category, but 
each ground instantiation of an axiom must respect the syntactic definition of SFM$_0$ given in 
Section \ref{cong-sec}; this means that we can write the axiom $x + (y + z) = (x + y) + z$, but 
it is invalid to instantiate it to $C + (a.\1 + b.\1) = (C + a.\1) + b.\1$ because these are not legal 
SFM$_0$ processes (the constant $C$ cannot be used as a summand).

\subsection{The Axioms}\label{set-axiom-sec}

The set of axioms is outlined in Table \ref{axiom-sfm0-tab}. 
The axioms {\bf A1-A4} are the usual axioms for choice \cite{Kleene,Salomaa,Kozen,Mil84}. 
The axiom schemata {\bf T1-T3} are those for prefixing \cite{Kleene,Salomaa,Kozen,Gor23nfa}, that are finitely many
as $A$ is finite. 
The conditional axioms {\bf R1-R2} are about process constants and are similar to those in \cite{Mil84,Gor19,Gor20ic}. 
Note that these conditional axioms are actually a finite collection of axioms, one for each constant 
definition: since the set $\cons$ of process constants is finite, the instances of {\bf R1-R2} are finitely many as well.

We call $B$ the set of axioms $\{${\bf A1, A2, A3, A4, R1, R2}$\}$ that is a sound and complete axiomatization for bisimilarity \cite{Mil84,Gor19,Gor20ic}, while we call
$W$ the set of axioms $\{${\bf A1, A2, A3, A4, T1, T2, T3, R1, R2}$\}$ that we will
prove to be a sound and complete axiomatization of language equivalence
for SFM$_0$ processes.
By the notation $E \vdash p = q$ we mean that there exists an equational deduction proof 
of the equality $p = q$, by using the axioms in $E$.

\begin{table}[t]
{\renewcommand{\arraystretch}{1.2}
\hrulefill\\[-.7cm]
			\begin{center}\fontsize{8pt}{0.1pt}

$\begin{array}{llrcll}
{\bf A1} &\; \;  \mbox{Associativity} &\; \;  x + (y + z) & = & (x + y) + z &\\
{\bf A2} &\; \;  \mbox{Commutativity} &\; \;  x + y & = & y + x& \\
{\bf A3} &\; \;  \mbox{Identity} &\; \;  x + \nil & = & x & \\
{\bf A4} &\; \;  \mbox{Idempotence} &\; \;  x + x & = & x & \\
\end{array}$

\hrulefill

$\begin{array}{llrcll}
{\bf T1} &\; \;  \mbox{Annihilation} &\; \;  a.\nil & = & \nil & \\
{\bf T2} &\; \;  \mbox{Distributivity} &\; \;  a.(x + y) & = & a.x + a.y& \\
{\bf T3} &\; \;  \mbox{$\epsilon$-Absorption} &\; \;  a.\epsilon.\1 & = & a.\1& \\
\end{array}$

\hrulefill

$\begin{array}{llrcllll}
{\bf R1} &\; \mbox{Unfolding} &\; \; \mbox{if $C \eqdef p \;$} & \mbox{then} &
\mbox{$C = p$} &\\
{\bf R2} & \;  \mbox{Folding} &  \mbox{if $C \eqdef p\{C/x\}  \; \wedge \;
q = p\{q/x\}$} & \mbox{then} & \mbox{$C = q$} & \\
\end{array}$

\hrulefill

\end{center}
}
\caption{Axioms for language equivalence}\label{axiom-sfm0-tab}
\end{table}

\begin{theorem}{\bf (Soundness)}\label{sound-th-sfm0}
For every $p, q \in  \mathcal{P}_{SFM_0}$, if $W \vdash p = q$, then $p \sim q$.
\proof The proof is by induction on the proof of $W \vdash p = q$. The 
thesis follows by observing that all the axioms in $W$ are sound  (e.g., {\bf T1-T3} by Proposition \ref{pref-alg-prop}) 
and that $ \sim$ is a congruence.
\fine
\end{theorem}

\subsection{Normal Forms and Unique Solutions}\label{det-nf-sec}

An SFM$_0$ process $p$ is a {\em normal form} if the predicate $nf(p)$ holds. This predicate stands for 
$nf(p, \emptyset)$, whose inductive definition is displayed in Table \ref{nf1-tab}. 
Examples of terms which are not in normal form are  $a.b.\1$, $b.\nil$ and $C \eqdef a.b.C + b.\epsilon.\1$.

\begin{table}[t]
{\renewcommand{\arraystretch}{1.6}
\hrulefill\\[-.6cm]
			\begin{center}\fontsize{8pt}{0.1pt}

$\begin{array}{cccccc}
\bigfrac{}{nf(\nil, I)} & \quad & \bigfrac{}{nf(\alpha.\1, I)} & \quad    & 
\bigfrac{nf(p, I \cup \{C\}) \quad C \eqdef p \quad C \not \in I}{nf(C, I)}\\

\bigfrac{nf(C, I)}{nf(a.C, I)}
& \quad & \bigfrac{nf(p, I) \quad nf(q, I)}{nf(p + q, I)} & \quad & \bigfrac{C \in I}{nf(C, I)} \\
\end{array}$

\hrulefill\\
\end{center}}
\caption{Normal form predicate}\label{nf1-tab}
\end{table}

Note that
if $C$ is a normal form, then its body (ignoring all the possible summands $\nil$ that can be absorbed via axioms {\bf A3})
is of the form $\sum_{i = 1}^{n} a_i.C_i + \sum_{j=1}^{m} \alpha_j.\1$
(assuming that $\sum_{i = 1}^{n} a_i.C_i$ is $\nil$ if $n = 0$, as well as that $\sum_{j=1}^{m} \alpha_j.\1$ is $\nil$ if $m = 0$) 
where, in turn, each $C_i$ is a normal form.
As a matter of fact, according to the construction in the proof of Theorem \ref{representability-gfa}, the processes representing 
reduced GFAs are in normal form.

Now we show that, for each SFM$_0$ process $p$, there exists a normal form $q$ such that $B \vdash p = q$. Since bisimilarity is
finer than language equivalence, this ensures that the normal form $q$ is language equivalent to $p$. 

\begin{proposition}\label{nf-prop}{\bf (Reduction to normal form)}
Given an SFM$_0$ process $p$, there exists a normal form $q$ such that $B \vdash p = q$.

\proof The proof is by induction on the structure of $p$, with the proviso to use a set $I$ of already scanned constants, 
in order to avoid looping on recursively defined constants, where $I$ is initially empty. 
We prove that for $(p, I)$ there exists a term $q$ such that $nf(q,I)$ holds
and $(B, I) \vdash p = q$, where this means that the equality $p = q$ can be derived by the axioms in $B$ when
each constant $C \in I$ is assumed to be equated to itself only. The thesis then follows by considering $(p, \emptyset)$.

The first base case is $(\nil, I)$; in such a case, $q = \nil$, because $nf(\nil, I)$ holds, and the thesis 
$(B, I) \vdash \nil = \nil$ follows by reflexivity.
The second base case is $(\alpha.\1, I)$ and it is analogous to the previous one.
The third base case  is $(C, I)$ when $C \in I$; in this case, $C$ is already a normal form, as  $nf(C, I)$ holds,
and $(B, I) \vdash C = C$ by reflexivity.

Case $(a.p, I)$: by induction, $(p, I)$ has an associated normal form $nf(q, I)$ such that $(B, I) \vdash p = q$;
hence, $(B, I) \vdash a.p = a.q$ by substitutivity. If $q$ is a constant, then $a.q$ is already a 
normal form. Otherwise,
take a new constant $C \eqdef q$, so that $nf(C, I)$ holds because $nf(q, I \cup \{C\})$ holds
(note that $C$ does not occur in $q$, so that this is the same as stating $nf(q, I)$, which holds by induction).
The required normal form is $a.C$. Indeed, $nf(a.C, I)$ holds because $nf(C, I)$ holds; moreover,
since $(B, I) \vdash p = q$ by induction and $(B, I) \vdash C = q$ by axiom {\bf R1}, we have that 
$(B, I) \vdash p = C$ by transitivity, so that
$(B, I) \vdash a.p = a.C$ by substitutivity.

Case $(p_1 + p_2, I)$: by induction, we can assume that for $(p_i, I)$ there exists a normal form $nf(q_i, I)$
such that $(B, I) \vdash p_i = q_i$ for $i = 1, 2$. Then, $nf(q_1 + q_2, I)$ holds and 
$(B, I) \vdash p_1 + p_2 = q_1 + q_2$ by substitutivity.

Case $(C, I)$, where $C \eqdef r\{C/x\}$ and  $C \not\in I$. 
By induction on $(r\{C/x\}, I \cup \{C\})$, we can assume that there exists 
a normal form $q\{C/x\}$ such that $nf(q\{C/x\}, I \cup \{C\})$ holds and
$(B, I \cup \{C\}) \vdash r\{C/x\} = q\{C/x\}$. Note that, by construction, if $(B, I \cup \{C\}) \vdash r\{C/x\} = q\{C/x\}$,
then also $(B, I) \vdash r\{C/x\} = q\{C/x\}$. Hence, as $(B, I) \vdash C =  r\{C/x\}$ by axiom {\bf R1}, it follows that
$(B, I) \vdash C = q\{C/x\}$ by transitivity. 
Then, we take a new constant $D \eqdef q\{D/x\}$ such that  $nf(D, I)$ because $nf(q\{D/x\}, I \cup \{D\})$ holds.
Hence, $(B, I) \vdash C = D$ by axiom {\bf R2}, where $D$ is
a normal form.
\fine
\end{proposition}

\begin{remark}{\bf (Normal forms as systems of equations)}\label{rem-normal}
As a matter of fact, we can restrict our attention only to 
normal forms defined by constants: if $p$ is a normal form and 
$p$ is not a constant, then take a new constant $D \eqdef p$, which is a normal form such that $B \vdash D = p$
by axiom {\bf R1}. For this reason, 
in the following we restrict our attention to
normal forms that can be defined by means of a {\em system of equations}:  a set 
$\widetilde{C} = \{C_1, C_2, \ldots, C_n\}$ of defined constants in normal form,
such that $\const{C_1} = \widetilde{C}$ and $\const{C_i} \subseteq \widetilde{C}$ for $i = 1, \ldots, n$.
The equations of the system $E(\widetilde{C})$ 
are of the following form (where function
$f(i,j)$ returns a value $k$ such that $1 \leq k \leq n$):\\

$\begin{array}{lcllcllcl}
C_1 & \eqdef &  \sum_{i = 1}^{m(1)} a_{1i}.C_{f(1,i)} + \sum_{j = 1}^{l(1)} \alpha_{1j}.\1\\
C_2 & \eqdef &  \sum_{i = 1}^{m(2)} a_{2i}.C_{f(2,i)} + \sum_{j = 1}^{l(2)} \alpha_{2j}.\1\\
\ldots\\
C_n & \eqdef & \sum_{i = 1}^{m(n)} a_{ni}.C_{f(n,i)} + \sum_{j = 1}^{l(n)} \alpha_{nj}.\1\\
\end{array}$\\

\noindent
where 
we assume $C_i \eqdef \nil$ in case $m(i) = 0 = l(i)$. We sometimes use the notation $body(C_h)$ to denote
the sumform $ \sum_{i = 1}^{m(h)} a_{hi}.C_{f(h,i)} + \sum_{j = 1}^{l(h)} \alpha_{hj}.\1$.
\fine
\end{remark}

The next theorem shows that every system of equations 
(not necessarily in normal form) has a unique solution up to provably equality.

\begin{theorem}{\bf (Unique solution for $W$)}\label{unique-w-th}
Let $\widetilde{X} = (x_1, x_2, \ldots, x_n)$ be a tuple of variables and let 
$\widetilde{p} = (p_1, p_2, \ldots, p_n)$ be a tuple of
open guarded SFM$_0$ terms, using the variables in $\widetilde{X}$. 
Let $\widetilde{C} = (C_1, C_2, \ldots, C_n)$ be a tuple of constants (not occurring in $\widetilde{p}$) 
used in the following system of equations
 
 $\begin{array}{lcllcllcl}
C_1 & \eqdef &  p_1\{\widetilde{C}/\widetilde{X}\} \\
C_2 & \eqdef & p_2\{\widetilde{C}/\widetilde{X}\} \\
\ldots\\
C_n & \eqdef & p_n\{\widetilde{C}/\widetilde{X}\} 
\end{array}$\\

\noindent
Then, for $i = 1, \ldots, n$, 
$W \vdash C_i = p_i\{\widetilde{C}/\widetilde{X}\}$.
Moreover, if the same property holds for $\widetilde{q} = (q_1, q_2, \ldots, q_n)$, i.e.,
for $i = 1, \ldots, n$ $W \vdash q_i = p_i\{\widetilde{q}/\widetilde{X}\} $, 
then 
$W \vdash C_i = q_i$.

\proof By induction on $n$. 

For $n = 1$, we have $C_1 \eqdef p_1\{C_1/x_1\}$,
and so the result $W \vdash C_1 = p_1\{C_1/x_1\}$
follows immediately using axiom {\bf R1}. 
This solution is unique: if $W \vdash q_1 = p_1\{q_1/x_1\}$, 
by axiom {\bf R2} we get $W \vdash C_1 = q_1$.
  
Now assume a tuple $\widetilde{p} = (p_1, p_2, \ldots, p_n)$ and the term $p_{n+1}$,
so that they are all open on $\widetilde{X} = (x_1, x_2, \ldots, x_n)$ and the additional $x_{n+1}$.
Assume, w.l.o.g., that $x_{n+1}$ occurs in $p_{n+1}$. 
First, define 
$C_{n+1} \eqdef p_{n+1}\{C_{n+1}/x_{n+1}\}$, so that $C_{n+1}$ is now open on $\widetilde{X}$.
Therefore, also for $i = 1, \ldots, n$, each $p_{i}\{C_{n+1}/x_{n+1}\}$ is now
open on $\widetilde{X}$.
The resulting 
system of equations is:

$\begin{array}{lcllcllcl}
C_1 & \eqdef &  p_1\{C_{n+1}/x_{n+1}\}\{\widetilde{C}/\widetilde{X}\}\\
C_2 & \eqdef & p_2\{C_{n+1}/x_{n+1}\}\{\widetilde{C}/\widetilde{X}\}\\
\ldots\\
C_n & \eqdef & p_n\{C_{n+1}/x_{n+1}\}\{\widetilde{C}/\widetilde{X}\}\\
\end{array}$

\noindent so that now we can use induction on $\widetilde{X}$ and 
$(p_{1}\{C_{n+1}/x_{n+1}\}, \ldots, p_{n}\{C_{n+1}/x_{n+1}\})$, to conclude that 
the tuple $\widetilde{C} = (C_1, C_2, \ldots, C_n)$ of closed constants
is such that for $i = 1, \ldots, n$:

$W \vdash C_i = (p_i\{C_{n+1}/x_{n+1}\})\{\widetilde{C}/\widetilde{X}\} = p_{i}\{\widetilde{C}/\widetilde{X}, 
C_{n+1}\{\widetilde{C}/\widetilde{X}\}/x_{n+1}\}$.

\noindent 
Note that above by $C_{n+1}\{\widetilde{C}/\widetilde{X}\}$ we have implicitly closed the definition of 
$C_{n+1}$ as
 
$C_{n+1} \eqdef (p_{n+1}\{C_{n+1}/x_{n+1}\})\{\widetilde{C}/\widetilde{X}\} = 
(p_{n+1}\{\widetilde{C}/\widetilde{X}\})\{C_{n+1}/x_{n+1}\}$,

\noindent
so that $W \vdash C_{n+1} = (p_{n+1}\{\widetilde{C}/\widetilde{X}\})\{C_{n+1}/x_{n+1}\}$ 
by axiom {\bf R1}. 

Unicity of the tuple $(\widetilde{C}, C_{n+1})$ can be proved by using axiom {\bf R2}.
Assume to have another solution tuple $(\widetilde{q}, q_{n+1})$. This means that, for $i = 1, \ldots, n+1$,

$W \vdash q_i = p_i\{\widetilde{q}/\widetilde{X}, q_{n+1}/x_{n+1}\}$.

\noindent 
By induction, we can assume, for $i = 1, \ldots, n$, that $W \vdash C_i = q_i$.

Since $W \vdash C_{n+1} = p_{n+1}\{\widetilde{C}/\widetilde{X}\}\{C_{n+1}/x_{n+1}\}$ 
by axiom {\bf R1},
by substitutivity we get $W \vdash C_{n+1} = p_{n+1}\{\widetilde{q}/\widetilde{X}\}\{C_{n+1}/x_{n+1}\}$.

Note that $p_{n+1}\{\widetilde{q}/\widetilde{X}\}$ is a term open on $x_{n+1}$ which is guarded.
Let $F$ be a constant defined as follows: $F \eqdef  p_{n+1}\{\widetilde{q}/\widetilde{X}\}\{F/x_{n+1}\}$.
Then, by axiom {\bf R2}, $C_{n+1} = F$.
Hence, since 
$W \vdash q_{n+1} = p_{n+1}\{\widetilde{q}/\widetilde{X}\}\{q_{n+1}/x_{n+1}\}$,
by axiom {\bf R2} we get $W \vdash F = q_{n+1}$; and so  the thesis  
$W \vdash C_{n+1} = q_{n+1}$ follows by transitivity.
\fine
\end{theorem}

\subsection{Saturated Normal Forms}\label{satur-nform-sec}

Now we want to show that it is possible to saturate a normal form, meaning that
if $C_h \Deriv{a} \1$, then $W \vdash C_h = body(C_h) + a.\1$.

\begin{example}
Consider the system of equations\\

$\begin{array}{rcl}
C_1 & \eqdef & a.C_1 +  b.C_2 + \epsilon.\1\\
C_2 & \eqdef & a.C_2 + \epsilon.\1\\
\end{array}$\\

\noindent
where the language recognized by $C_1$ is $a^* | a^* b a^*$. We can get an equivalent system of
equations by adding terminal summands as follows: whenever a summand $a.C_i$ is present in the body of
$C_j$ and $C_i$ has 
an $\epsilon.\1$ summand, add to the definition of $C_j$ also a summand $a.\1$.
Applying this construction to the system of equations above, we get\\

$\begin{array}{rcl}
C'_1 & \eqdef & a.C'_1 +  b.C'_2 + a.\1 + b.\1 + \epsilon.\1\\
C'_2 & \eqdef & a.C'_2 + a.\1 + \epsilon.\1\\
\end{array}$\\

\noindent
such that $W \vdash C_i = C'_i$ for $i = 1, 2$.
\fine
\end{example}

\begin{lemma}{\bf (Saturation Lemma)}\label{satur-lemma}
Given a normal form

$\begin{array}{lcllcllcl}
C_1 & \eqdef &  \sum_{i = 1}^{m(1)} a_{1i}.C_{f(1,i)} + \sum_{j = 1}^{l(1)} \alpha_{1j}.\1\\
C_2 & \eqdef &  \sum_{i = 1}^{m(2)} a_{2i}.C_{f(2,i)} + \sum_{j = 1}^{l(2)} \alpha_{2j}.\1\\
\ldots\\
C_n & \eqdef & \sum_{i = 1}^{m(n)} a_{ni}.C_{f(n,i)} + \sum_{j = 1}^{l(n)} \alpha_{nj}.\1\\
\end{array}$\\

\noindent
if $C_h \Deriv{a} \1$, then $W \vdash C_h = \sum_{i = 1}^{m(h)} a_{hi}.C_{f(h,i)} + \sum_{j = 1}^{l(h)} \alpha_{hj}.\1 + a.\1$.

\proof
If if $C_h \Deriv{a} \1$, then either $C_h \deriv{a} \1$, or $C_h \deriv{a} C_k \deriv{\epsilon} \1$ for some $1 \leq k \leq n$.
In the former case, the thesis follows by axiom {\bf A4}, as $a.\1$ is already a summand in the body of $C_h$.
In the latter case, first note that,
by axiom {\bf R1}, we have that $C_h = \sum_{i = 1}^{m(h)} a_{hi}.C_{f(h,i)} + \sum_{j = 1}^{l(h)} \alpha_{hj}.\1$.
This means that there exists an index $1 \leq i \leq m(h)$ such that $a_{hi} = a$ and $f(h,i) = k$.
By axiom {\bf R1}, $C_k = \sum_{i = 1}^{m(k)} a_{ki}.C_{f(k,i)} + \sum_{j = 1}^{l(k)} \alpha_{kj}.\1$.
By substitutivity and {\bf A4}, 

$C_h = \sum_{i = 1}^{m(h)} a_{hi}.C_{f(h,i)} + \sum_{j = 1}^{l(h)} \alpha_{hj}.\1 + a.body(C_k)$.

\noindent By axiom {\bf T2}, $a.body(C_k) = \sum_{i = 1}^{m(k)} a.a_{ki}.C_{f(k,i)} + \sum_{j = 1}^{l(k)} a.\alpha_{kj}.\1$.
One of the summand $\alpha_{kj}$ must be $\epsilon$, so that by axiom {\bf T3}, we have
$ a.\epsilon.\1 = a.\1$. 

By reversing the steps above, we end up into the required equality

$C_h = \sum_{i = 1}^{m(h)} a_{hi}.C_{f(h,i)} + \sum_{j = 1}^{l(h)} \alpha_{hj}.\1 + a.\1$.
\fine
\end{lemma}

\begin{proposition}\label{satur-prop}{\bf (Reduction to saturated normal form)}
Let $\widetilde{X} = (x_1, x_2, \ldots, x_n)$ be a tuple of variables. Let 
$\widetilde{p} = (p_1, p_2, \ldots, p_n)$ be a tuple of
open guarded SFM$_0$ terms, using the variables in $\widetilde{X}$, of the form 
$p_h = \sum_{i = 1}^{m(h)} a_{hi}.x_{f(h,i)} + \sum_{j = 1}^{l(h)} \alpha_{hj}.\1$
for $h = 1, \ldots, n$ and where $f$ is such that $1 \leq f(h,i) \leq n$.
Let $\widetilde{C} = (C_1, C_2, \ldots, C_n)$ be the tuple of constants 
 
 $\begin{array}{lcllcllcl}
C_1 & \eqdef &  p_1\{\widetilde{C}/\widetilde{X}\}\\
C_2 & \eqdef & p_2\{\widetilde{C}/\widetilde{X}\}\\
\ldots\\
C_n & \eqdef & p_n\{\widetilde{C}/\widetilde{X}\}
\end{array}$\\

\noindent
Then, there exist a tuple $\widetilde{q} = (q_1, q_2, \ldots, q_n)$ of
open guarded SFM$_0$ terms, using the variables in $\widetilde{X}$, and a tuple 
of constants $\widetilde{D} = (D_1, D_2, \ldots, D_n)$  such that the system of equations in normal form
 
 $\begin{array}{lcllcllcl}
D_1 & \eqdef &  q_1\{\widetilde{D}/\widetilde{X}\}\\
D_2 & \eqdef & q_2\{\widetilde{D}/\widetilde{X}\}\\
\ldots\\
D_n & \eqdef & q_n\{\widetilde{C}/\widetilde{X}\}
\end{array}$\\

\noindent
is saturated and 
$W_g \vdash C_i = D_i$ for $i = 1, \ldots, n$.

\proof By induction on $n$. For $n = 1$, we have that $C_1 \eqdef p_1\{C_1/x_1\}$,
where $p_1$ must be of the form
$\sum_{i = 1}^{m(1)} a_{1i}.x_{1} + \sum_{j = 1}^{l(1)} \alpha_{1j}.\1$.
If none of the $\alpha_{1j}$ is $\epsilon$, 
then $C_1$ is already saturated.
Otherwise, we have to saturate it, as explained in Lemma \ref{satur-lemma}, by defining
a new constant 

$D_1 \eqdef \sum_{i = 1}^{m(1)} a_{1i}.D_{1} + \sum_{j = 1}^{l(1)} \alpha_{1j}.\1
+ \sum_{i = 1}^{m(1)} a_{1i}.\1$,

\noindent
so that $W \vdash C_1 = D_1$.

Now assume a tuple $\widetilde{p} = (p_1, p_2, \ldots, p_n)$ and the term $p_{n+1}$,
so that they are all open on $\widetilde{X} = (x_1, x_2, \ldots, x_n)$ and the additional $x_{n+1}$.
Assume, w.l.o.g., that $x_{n+1}$ occurs in $p_{n+1}$. 
First, define 
$C_{n+1} \eqdef p_{n+1}\{C_{n+1}/x_{n+1}\}$, so that $C_{n+1}$ is now open on $\widetilde{X}$.
Therefore, also for $i = 1, \ldots, n$, each $p_{i}\{C_{n+1}/x_{n+1}\}$ is now
open on $\widetilde{X}$.
The resulting system of equations in normal form of size $n$ is:

$\begin{array}{lcllcllcl}
C_1 & \eqdef &  p_1\{C_{n+1}/x_{n+1}\}\{\widetilde{C}/\widetilde{X}\}\\
C_2 & \eqdef & p_2\{C_{n+1}/x_{n+1}\}\{\widetilde{C}/\widetilde{X}\}\\
\ldots\\
C_n & \eqdef & p_n\{C_{n+1}/x_{n+1}\}\{\widetilde{C}/\widetilde{X}\}\\
\end{array}$

\noindent 
so that, by induction, we can conclude that
there exist a tuple $\widetilde{q} = (q_1, q_2, \ldots, q_n)$ of terms and a tuple of constants $\widetilde{D} = (D_1, D_2, \ldots, D_n)$
such that 

$\begin{array}{lcllcllcl}
D_1 & \eqdef &  q_1\{C_{n+1}/x_{n+1}\}\{\widetilde{D}/\widetilde{X}\}\\
D_2 & \eqdef & q_2\{C_{n+1}/x_{n+1}\}\{\widetilde{D}/\widetilde{X}\}\\
\ldots\\
D_n & \eqdef & q_n\{C_{n+1}/x_{n+1}\}\{\widetilde{D}/\widetilde{X}\}
\end{array}$\\

\noindent 
is saturated and $W \vdash C_i = D_i$ for $i = 1, \ldots, n$.

Note that 

$q_i\{C_{n+1}/x_{n+1}\}\{\widetilde{D}/\widetilde{X}\} = q_{i}\{\widetilde{D}/\widetilde{X}, C_{n+1}\{\widetilde{D}/\widetilde{X}\}/x_{n+1}\}$

\noindent
so that by $C_{n+1}\{\widetilde{D}/\widetilde{X}\}$ we have implicitly closed the definition of 
$C_{n+1}$ as
 
$C_{n+1} \eqdef p_{n+1}\{C_{n+1}/x_{n+1}\}\{\widetilde{D}/\widetilde{X}\} = 
p_{n+1}\{\widetilde{D}/\widetilde{X}\}\{C_{n+1}/x_{n+1}\}$.

Note that inside the term $p_{n+1}\{\widetilde{D}/\widetilde{X}\}$ there can be
a summand $\epsilon.\1$. In such a case, for each term $q_i$ and for each
summand of the form $a_j.x_{n+1}$
inside term $q_i$, we have to add to the definition of $q_i$ also the summand $a_j.\1$,
resulting in the new saturated term $q_i'$.
Note that, by axioms $\{${\bf T3,R1,A4}$\}$ (as explained in Lemma \ref{satur-lemma}), we have

$W \vdash q_{i}\{\widetilde{D}/\widetilde{X}, C_{n+1}\{\widetilde{D}/\widetilde{X}\}/x_{n+1}\}
= q'_{i}\{\widetilde{D}/\widetilde{X}, C_{n+1}\{\widetilde{D}/\widetilde{X}\}/x_{n+1}\}
$.

We define the new system of equations in saturated normal form

$\begin{array}{lcllcllcl}
E_1 & \eqdef &  q'_1\{E_{n+1}/x_{n+1}\}\{\widetilde{D}/\widetilde{X}\}\\
E_2 & \eqdef & q'_2\{E_{n+1}/x_{n+1}\}\{\widetilde{D}/\widetilde{X}\}\\
\ldots\\
E_n & \eqdef & q'_n\{E_{n+1}/x_{n+1}\}\{\widetilde{D}/\widetilde{X}\}\\
\end{array}$\\

\noindent
such that $W \vdash D_i  = E_i$ for $i = 1, \ldots, n$ by recursion congruence.

If inside the term $p_{n+1}\{\widetilde{D}/\widetilde{X}\}$ there is
a summand $\epsilon.\1$, then we have to saturate $p_{n+1}\{\widetilde{D}/\widetilde{X}\}$,
yielding $p'_{n+1}\{\widetilde{D}/\widetilde{X}\}$, such that 

$W \vdash p_{n+1}\{\widetilde{D}/\widetilde{X}\}
= p'_{n+1}\{\widetilde{D}/\widetilde{X}\}
$.

Note also that inside the term $p_{n+1}\{\widetilde{D}/\widetilde{X}\}$ there can be
summands of the form $a_j.C_h$; in case $\epsilon.\1$ is a summand for $C_h$,
for each $a_j.C_h$, we have to add to the term $p'_{n+1}\{\widetilde{D}/\widetilde{X}\}$
also the summand $a_j.\1$, resulting in the new term $q'_{n+1}\{\widetilde{D}/\widetilde{X}\}$.
Note that, by axioms $\{${\bf T3,R1,A4}$\}$, we have

$W \vdash p'_{n+1}\{\widetilde{D}/\widetilde{X}\}
= q'_{n+1}\{\widetilde{D}/\widetilde{X}\}
$.

Now define a new constant $E_{n+1} \eqdef q'_{n+1}\{\widetilde{D}/\widetilde{X}\}\{E_{n+1}/x_{n+1}\}$ so that
$W \vdash C_{n+1} = E_{n+1}$ by recursion congruence.

Moreover, by axiom {\bf R1}, we have that 
$W \vdash E_{n+1} = q'_{n+1}\{E_{n+1}/x_{n+1}\}\{\widetilde{D}/\widetilde{X}\}$. Since we have that
 $W \vdash D_i  = E_i$ for $i = 1, \ldots, n$,  by recursion congruence
 we also get 
 
 $W \vdash E_{n+1} = q'_{n+1}\{E_{n+1}/x_{n+1}\}\{\widetilde{E}/\widetilde{X}\}$.

Finally, 
we define a new system of equations in saturated normal form

$\begin{array}{lcllcllcl}
F_1 & \eqdef &  q'_1\{F_{n+1}/x_{n+1}\}\{\widetilde{F}/\widetilde{X}\}\\
F_2 & \eqdef & q'_2\{F_{n+1}/x_{n+1}\}\{\widetilde{F}/\widetilde{X}\}\\
\ldots\\
F_n & \eqdef & q'_n\{F_{n+1}/x_{n+1}\}\{\widetilde{F}/\widetilde{X}\}\\
F_{n+1} & \eqdef & q'_{n+1}\{F_{n+1}/x_{n+1}\}\{\widetilde{F}/\widetilde{X}\}
\end{array}$\\

\noindent
such that $W \vdash E_i  = F_i$ for $i = 1, \ldots, n+1$ by Theorem \ref{unique-w-th}. 
The thesis, $W \vdash C_i = F_i$,  for $i = 1, \ldots, n$,
follows by transitivity.
\fine
\end{proposition}

\subsection{Semi-deterministic Normal Form}\label{semi-det-nform-sec}

Given an alphabet $A = \{a_1, \ldots, a_k\}$, we say that a
normal form is semi-deterministic if it is of the form\\

$\begin{array}{lcllcllcl}
C_1 & \eqdef &  \sum_{i = 1}^{k} a_{i}.C_{f(1,i)} + \sum_{j = 1}^{l(1)} \alpha_{1j}.\1\\
C_2 & \eqdef &  \sum_{i = 1}^{k} a_{i}.C_{f(2,i)} + \sum_{j = 1}^{l(2)} \alpha_{2j}.\1\\
\ldots\\
C_n & \eqdef & \sum_{i = 1}^{k} a_{i}.C_{f(n,i)} + \sum_{j = 1}^{l(n)} \alpha_{nj}.\1\\
\end{array}$\\

\noindent
where function
$f(i,j)$ returns a value $k$ such that $1 \leq k \leq n$ and, for $h = 1, \ldots, n$, the 
multiset $\{\alpha_{h1}, \alpha_{h2}, \ldots, \alpha_{hl(h)}\}$ is a actually a set. Therefore,
the body of each constant contains exactly one {\em non-terminal} summand (i.e., ending with a constant) 
for each symbol $a_i \in A$ and at most
one {\em terminal} summand (i.e., ending with $\1$) for each symbol $\alpha_i \in A \cup \{\epsilon\}$. Of course,
the GFA associated to $C_1$ above, 
according to the denotational semantics in Table \ref{den-gfa-sfm-epsilon}, is semi-deterministic 
and has exactly $A$ as its alphabet. Conversely, according to the construction in the proof of Theorem \ref{representability-gfa}, 
the process terms representing 
reduced DGFAs are in semi-deterministic normal form.

Note that the concept of semi-determinism heavily depends on the chosen alphabet. 
For instance, according to the denotational semantics in Table \ref{den-gfa-sfm-epsilon}, the GFA for $C \eqdef \nil$ is semi-deterministic, because its alphabet is empty. 
However, it is not semi-deterministic if we consider 
a nonempty alphabet, say $A = \{a\}$,
because there is not an arc for $a$. In such a case, a DGFA w.r.t. $A$, language equivalent to $C \eqdef \nil$, could be the one associated to
$E \eqdef a.E$. Indeed, it is possible to prove that $W \vdash C = E$ by means of axioms {\bf T1, R1, R2}; in fact, 
by axiom {\bf T1}, we have $W \vdash \nil = a.\nil$; then, by axiom {\bf R2}, from $E \eqdef a.E$ and 
$W \vdash \nil = a.\nil$, we get $W \vdash E = \nil$; then by axiom {\bf R1}, we have $W \vdash C = \nil$, so that
$W_g \vdash C = E$ follows by transitivity.

Similarly, the GFA for $C \eqdef \epsilon.\1$ is semi-deterministic, because its alphabet is empty. 
However, it is not semi-deterministic if we consider 
the nonempty alphabet $A = \{a\}$. In such a case, a DGFA w.r.t. $A$, language equivalent to $C \eqdef \epsilon.\1$, could be the one associated to
$D \eqdef a.E + \epsilon.\1$, where $E \eqdef a.E$ denotes a sink, i.e., an error state. 
We can prove that $W \vdash C = D$ as follows.
First, we already know that $W \vdash E = \nil$; so, by substitutivity and {\bf R1}, we have 
$W \vdash D = a.\nil + \epsilon.\1$; 
then by {\bf T1},
$W \vdash D = \nil + \epsilon.\1$; then by axioms {\bf A2,A3}, we have $W \vdash D = \epsilon.\1$, so that $W \vdash C = D$ follows easily.

Now we prove that, given a normal form, it is possible to construct a provably equivalent semi-deterministic normal form.
The idea behind the proof, that follows the classic Rabin-Scott {\em subset construction} \cite{RS59}, is illustrated by the following example.

\begin{example}\label{ex-det}
Let us consider the following normal form

$\begin{array}{rcl}
C_1 & \eqdef & a.C_1 +  a.C_2 +  a.C_1 + \epsilon.\1\\
C_2 & \eqdef & a.C_2 + \epsilon.\1\\
\end{array}$\\

\noindent
whose alphabet $A_C$ is $\{a\}$. Let us take an alphabet $A = \{a, b\}$, w.r.t. which we want to define an equivalent 
semi-deterministic normal form. First, we define a collection of set-indexed 
constants $B_I$, where $I \subseteq \{1, 2\}$, as $B_I \eqdef \sum_{i \in I} body(C_i)$. More explicitly,

$\begin{array}{rcl}
B_{\{1\}} & \eqdef & a.C_1 +  a.C_2 +  a.C_1 + \epsilon.\1\\
B_{\{2\}} & \eqdef & a.C_2 + \epsilon.\1\\
B_{\{1,2\}} & \eqdef & a.C_1 +  a.C_2 +  a.C_1 + \epsilon.\1 + a.C_2 + \epsilon.\1\\
B_\emptyset & \eqdef & \nil\\
\end{array}$\\

\noindent
Now, for each of them, we prove that it can be equated to a semi-deterministic definition. First, we can prove that, 
since $B_\emptyset = \nil$ by axiom {\bf R1}, also
$W \vdash B_\emptyset = a.B_{\emptyset} + b.B_{\emptyset}$ by axioms {\bf T1,A3}.

Now, by axiom {\bf R1}, we know that $W \vdash B_{\{1\}} = a.C_1 +  a.C_2 +  a.C_1 + \epsilon.\1$, so 
that $W \vdash C_1 = B_{\{1\}}$. Then, by axioms {\bf A1-A4,T2,R1},
we can derive $W \vdash B_{\{1\}} = a.(body(C_1) + body(C_2)) +  \epsilon.\1$. Then, we can also derive 
$W \vdash B_{\{1\}} = a.B_{\{1,2\}} + b.B_\emptyset + \epsilon.\1$ by axioms {\bf R1,T1,A3}. 

In the same way, we can prove that $W \vdash B_{\{2\}} = a.B_{\{2\}} + b.B_\emptyset + \epsilon.\1$ and, moreover, that
$W \vdash B_{\{1,2\}} = a.B_{\{1,2\}} + b.B_\emptyset + \epsilon.\1$, where the duplicated summand $\epsilon.\1$ is taken only once
(by axiom {\bf A4}).

Finally, we define a new semi-deterministic system of equations

$\begin{array}{rcl}
D_{\{1\}} & \eqdef & a.D_{\{1,2\}} + b.D_{\emptyset} + \epsilon.\1\\
D_{\{2\}} & \eqdef & a.D_{\{2\}} + b.D_{\emptyset} + \epsilon.\1\\
D_{\{1,2\}} & \eqdef &  a.D_{\{1,2\}} + b.D_{\emptyset} + \epsilon.\1\\
D_{\emptyset} & \eqdef & a.D_{\emptyset} + b.D_{\emptyset}\\
\end{array}$\\

\noindent
where $D_{\{2\}}$ is redundant as $D_{\{2\}} \not \in \const{D_{\{1\}}}$, such that, by Theorem \ref{unique-w-th}, 
we have $W \vdash B_I = D_I$ for $I = \{1\}, \{2\}, \{1,2\}, \emptyset$.
So, $W \vdash C_1 = D_{\{1\}}$ by transitivity.
\fine
\end{example}

\begin{proposition}\label{det-red-prop}{\bf (Reduction to semi-deterministic normal form)}
Given the normal form
 
 $\begin{array}{lcllcllcl}
C_1 & \eqdef &  \sum_{i = 1}^{m(1)} a_{1i}.C_{f(1,i)} + \sum_{j = 1}^{l(1)} \alpha_{1j}.\1\\
C_2 & \eqdef &  \sum_{i = 1}^{m(2)} a_{2i}.C_{f(2,i)} + \sum_{j = 1}^{l(2)} \alpha_{2j}.\1\\
\ldots\\
C_n & \eqdef & \sum_{i = 1}^{m(n)} a_{ni}.C_{f(n,i)} + \sum_{j = 1}^{l(n)} \alpha_{nj}.\1\\
\end{array}$\\

\noindent 
such that its alphabet is $A_{C}$ and $1 \leq f(i,j) \leq n$,
it is possible to construct a semi-deterministic normal form w.r.t. a chosen $A = \{a_1, a_2, \ldots, a_k\}$,
with $A_C \subseteq A$, of the form 

$\begin{array}{lcllcllcl}
D_1 & \eqdef &  \sum_{i = 1}^{k} a_{i}.D_{g(1,i)} + \sum_{j = 1}^{k(1)} \alpha_{j}^1.\1\\
D_2 & \eqdef &  \sum_{i = 1}^{k} a_{i}.D_{g(2,i)} + \sum_{j = 1}^{k(2)} \alpha_{j}^2.\1\\
\ldots\\
D_m & \eqdef & \sum_{i = 1}^{k} a_{i}.D_{g(m,i)} + \sum_{j = 1}^{k(m)} \alpha_{j}^m.\1\\
\end{array}$\\

\noindent
where $m \geq n$ and $1 \leq g(i,j) \leq m$, such that $W \vdash C_1 = D_1$.

\proof 
We define
a set of set-indexed constants $B_I \eqdef \sum_{i \in I} body(C_i)$, one for each $I \subseteq \{1, \ldots, n\}$.
Note that $W \vdash C_1 = B_{\{1\}}$ by axiom {\bf R1}.
Now for each of these constants $B_I$ we want to show that it can be equated to a semi-deterministic 
definition, i.e., a definition where there is exactly one non-terminal summand per symbol in $A$ and at most one terminal summand
per symbol in $A \cup \{\epsilon\}$. 
First, let $In_n(B_I)$ denote the set of initial non-terminal symbols of $B_I$, i.e.,
$\cup_{i \in I} \{a_{i\,1}, \ldots, a_{i\, m(i)}\}$.
Second, let $In_f(B_I)$ denote the set of initial terminal symbols of $B_I$, i.e.,
$\cup_{i \in I} \{\alpha_{i\,1}, \ldots, \alpha_{i\, l(i)}\}$.
Third, we assume that $\alpha_h = a_h$, for $i = 1, \ldots, k$, 
and  $\alpha_{k+1} = \epsilon$.

Let us define, for $h = 1, \ldots, k$, the set $I_h = \{f(i,j) \mid a_{ij} = a_h, a_{ij} \in In_n(B_I)\}$.
Then, by using the axioms {\bf A1-A4,R1,T2}  and substitutivity, we can prove that

$W \vdash B_I = \sum_{a_{h}\in In_n(B_I)} a_{h}.(\sum_{i \in I_h} body(C_i)) + \sum_{\alpha_{h}\in In_f(B_I)} \alpha_h.\1$.

\noindent
The next step is to recognize that $\sum_{i \in I_h} body(C_i)$ can be replaced by $B_{I_h}$ by axiom {\bf R1}, so that
$W \vdash B_I = \sum_{a_{h}\in In_n(B_I)} a_{h}.B_{I_h} + \sum_{\alpha_{h}\in In_f(B_I)} \alpha_h.\1$.

The next step is to add non-terminal summands related to those symbols in $A$ not in $In_n(B_I)$. 
First note that $B_\emptyset \eqdef \nil$ 
can be equated, by means of axioms {\bf R1,T1,A3}, to 

$W \vdash B_\emptyset = \sum_{a_i \in A} a_i.B_\emptyset$, 

\noindent 
which is a semi-deterministic definition for the error state $B_\emptyset$. Therefore, by axioms {\bf T1,A3} we have

$W \vdash B_I = \sum_{a_{h}\in In_n(B_I)} a_{h}.B_{I_h} + \sum_{a_h \not\in In_n(B_I)} a_h.B_\emptyset 
+ \sum_{\alpha_{h}\in In_f(B_I)} \alpha_h.\1$

\noindent 
which is a semi-deterministic definition for constant $B_I$. Finally, we can define a semi-deterministic system of equations 
of the form

$D_I \eqdef \sum_{a_{h}\in In_n(B_I)} a_{h}.D_{I_h} + \sum_{a_h \not\in In_n(B_I)} a_h.D_\emptyset 
+ \sum_{\alpha_{h}\in In_f(B_I)} \alpha_h.\1$

\noindent
for each $I \subseteq \{1, \ldots, n\}$, such that, by Theorem \ref{unique-w-th}, we have $W_g \vdash B_I = D_I$. 
So, the thesis $W \vdash C_1 = D_{\{1\}}$ follows by transitivity.
\fine
\end{proposition}

\begin{corollary}\label{satur-det-cor}{\bf (Reduction to saturated semi-deterministic normal form)}
Given the saturated normal form
 
 $\begin{array}{lcllcllcl}
C_1 & \eqdef &  \sum_{i = 1}^{m(1)} a_{1i}.C_{f(1,i)} + \sum_{j = 1}^{l(1)} \alpha_{1j}.\1\\
C_2 & \eqdef &  \sum_{i = 1}^{m(2)} a_{2i}.C_{f(2,i)} + \sum_{j = 1}^{l(2)} \alpha_{2j}.\1\\
\ldots\\
C_n & \eqdef & \sum_{i = 1}^{m(n)} a_{ni}.C_{f(n,i)} + \sum_{j = 1}^{l(n)} \alpha_{nj}.\1\\
\end{array}$\\

\noindent 
such that its alphabet is $A_{C}$ and $1 \leq f(i,j) \leq n$,
it is possible to construct a semi-deterministic saturated normal form w.r.t. a chosen $A = \{a_1, a_2, \ldots, a_k\}$,
with $A_C \subseteq A$, of the form 

$\begin{array}{lcllcllcl}
D_1 & \eqdef &  \sum_{i = 1}^{k} a_{i}.D_{g(1,i)} + \sum_{j = 1}^{k(1)} \alpha_{j}^1.\1\\
D_2 & \eqdef &  \sum_{i = 1}^{k} a_{i}.D_{g(2,i)} + \sum_{j = 1}^{k(2)} \alpha_{j}^2.\1\\
\ldots\\
D_m & \eqdef & \sum_{i = 1}^{k} a_{i}.D_{g(m,i)} + \sum_{j = 1}^{k(m)} \alpha_{j}^m.\1\\
\end{array}$\\

\noindent
where $m \geq n$ and $1 \leq g(i,j) \leq m$, such that $W \vdash C_1 = D_1$.

\proof
The proof of Proposition \ref{det-red-prop} ensures that if we start from saturated normal forms,
then we get semi-deterministic saturated normal forms.
\fine
\end{corollary}

\begin{remark}{\bf (Removing $\epsilon$-labeled transitions)}\label{espilon-free-satur-semidet-rem}
Given the semi-deterministic saturated normal form
 
 $\begin{array}{lcllcllcl}
C_1 & \eqdef &  \sum_{i = 1}^{m(1)} a_{1i}.C_{f(1,i)} + \sum_{j = 1}^{l(1)} \alpha_{1j}.\1\\
C_2 & \eqdef &  \sum_{i = 1}^{m(2)} a_{2i}.C_{f(2,i)} + \sum_{j = 1}^{l(2)} \alpha_{2j}.\1\\
\ldots\\
C_n & \eqdef & \sum_{i = 1}^{m(n)} a_{ni}.C_{f(n,i)} + \sum_{j = 1}^{l(n)} \alpha_{nj}.\1\\
\end{array}$\\

\noindent 
it is possible to construct a semi-deterministic saturated {\em $\epsilon$-free}  normal form 
by simply removing all the possible $\epsilon.\1$ summands from the definitions al all the constants,
except for $C_1$ (cf. also Remark \ref{satur-epsilon-equiv}). The resulting normal form 

$\begin{array}{lcllcllcl}
D_1 & \eqdef &  \sum_{i = 1}^{k} a_{i}.D_{g(1,i)} + \sum_{j = 1}^{k(1)} \alpha_{j}^1.\1\\
D_2 & \eqdef &  \sum_{i = 1}^{k} a_{i}.D_{g(2,i)} + \sum_{j = 1}^{k(2)} a_{j}^2.\1\\
\ldots\\
D_n & \eqdef & \sum_{i = 1}^{k} a_{i}.D_{g(n,i)} + \sum_{j = 1}^{k(n)} a_{j}^n.\1\\
\end{array}$\\

\noindent
besides being semi-deterministic and saturated, is also $\epsilon$-free and $W \vdash C_1 = D_1$.
\fine
\end{remark}

\subsection{Completeness}\label{compl-obg-sec}

Corollary \ref{satur-det-cor} and Remark \ref{espilon-free-satur-semidet-rem} are at the basis of the lemma below, 
which, in turn, is crucial for the proof of completeness 
that follows. The proof technique of the lemma below is inspired to the proof
of Theorem 5.10 in \cite{Mil84}, in turn inspired to the proof of Theorem 2 in \cite{Salomaa}.

\begin{lemma}{\bf (Completeness for saturated normal forms)}\label{det-nf-compl}
For every $p, p' \in  \mathcal{P}_{SFM_0}$ saturated normal forms, 
if $p \sim p'$, then $W \vdash p = p'$.

\proof Let $A_p$ the alphabet of $p$ and $A_{p'}$ the alphabet of $p'$. Take $A = A_{p} \cup A_{p'}$.
By Corollary \ref{satur-det-cor}, there exist $p_1$ and $p_1'$ semi-deterministic saturated normal forms w.r.t. $A$,
such that $W \vdash p = p_1$ and $W \vdash p' = p_1'$. 
By Theorem \ref{sound-th-sfm0}, we have $p \sim p_1$
and $p' \sim p_1'$ and so, by transitivity, also $p_1 \sim p_1'$.
By Remark \ref{espilon-free-satur-semidet-rem}, there exists $p_2$ and $p_2'$ semi-deterministic saturated 
$\epsilon$-free normal forms w.r.t. $A$, such that $W \vdash p_1 = p_2$ and $W \vdash p'_1 = p_2'$. 
By Theorem \ref{sound-th-sfm0}, we have $p_1 \sim p_2$
and $p_1' \sim p_2'$ and so, by transitivity, also $p_2 \sim p_2'$.

Let $A = \{a_1, \ldots, a_k\}$. We can assume that $p_2$
is the saturated semi-deterministic $\epsilon$-free system of equations: 

$\begin{array}{lcllcllcl}
C_1 & \eqdef &  \sum_{i = 1}^{k} a_{i}.C_{f(1,i)} + \sum_{j = 1}^{l(1)} \alpha_{1j}.\1\\
C_2 & \eqdef &  \sum_{i = 1}^{k} a_{i}.C_{f(2,i)} + \sum_{j = 1}^{l(2)} \alpha_{2j}.\1\\
\ldots\\
C_n & \eqdef & \sum_{i = 1}^{k} a_{i}.C_{f(n,i)} + \sum_{j = 1}^{l(n)} \alpha_{nj}.\1\\
\end{array}$\\

\noindent 
so that $p_2 = C_1$. For each $h = 1, \ldots, n$,
we get
$W \vdash  C_h = body(C_h)$, by axiom {\bf R1}, where by $body(C_h)$ we denote 
$\sum_{i = 1}^{k} a_{i}.C_{f(h,i)} + \sum_{j = 1}^{l(h)} \alpha_{hj}.\1$.

Similarly, we can assume that $p_2'$ is the saturated semi-deterministic $\epsilon$-free system of 
equations: 

$\begin{array}{lcllcllcl}
C'_1 & \eqdef &  \sum_{i = 1}^{k} a_{i}.C'_{f'(1,i)} + \sum_{j = 1}^{l'(1)} \alpha'_{1j}.\1\\
C'_2 & \eqdef &  \sum_{i = 1}^{k} a_{i}.C'_{f'(2,i)} + \sum_{j = 1}^{l'(2)} \alpha'_{2j}.\1\\
\ldots\\
C'_{n'} & \eqdef & \sum_{i = 1}^{k} a_{i}.C'_{f'(n',i)} + \sum_{j = 1}^{l'(n')} \alpha'_{n'j}.\1\\
\end{array}$\\

\noindent 
so that $p_2' = C'_1$. For each $h' = 1, \ldots, n'$, 
we get
$W \vdash  C'_{h'} = body(C'_{h'})$, by axiom {\bf R1}, where by $body(C'_{h'})$ we denote 
$\sum_{i = 1}^{k} a_{i}.C'_{f'(h',i)} + \sum_{j = 1}^{l'(h')} \alpha'_{h'j}.\1$.

Moreover, as $p_2 \sim p_2'$, we have $C_1 \sim C'_1$. 

Now, let $H = \{(h, h') \mid $  
$ C_h \sim C'_{h'}\}$. Of course, note that $(1,1) \in H$.

Moreover, for $(h, h') \in H$, since  $C_h$ and $C'_{h'}$ are language equivalent, saturated, $\epsilon$-free and 
semi-deterministic (hence, 
bisimulation equivalent, cf. Remark \ref{lang=bis-rem}), the following hold:
\begin{itemize}
\item for each $i = 1, \ldots, k$, we have $(f(h,i), f'(h',i)) \in H$, and
\item a terminal summand $\alpha.\1$ belongs to $body(C_h)$ iff it belongs to $body(C'_{h'})$, i.e., 
{\bf \{A1-A2\}} $\vdash  \sum_{j = 1}^{l(h)} \alpha_{hj}.\1 = \sum_{j = 1}^{l'(h')} \alpha'_{h'j}.\1$.
\end{itemize}

\noindent
Now, for each $(h, h') \in H$, let us consider the open term 

$
t_{hh'} = \sum_{i = 1}^{k}  a_{i}.x_{f(h,i),f'(h',i)} + \sum_{j = 1}^{l(h)} \alpha_{hj}.\1
$

\noindent
By Theorem \ref{unique-w-th}, for each $(h, h') \in H$,  there exists a constant $D_{hh'}$ such that $W \vdash D_{hh'} 
= t_{hh'}\{\widetilde{D}/\widetilde{X}\} $, where $\widetilde{D}$ denotes the tuple of constants $D_{hh'}$ 
for each $(h, h') \in H$, and $\widetilde{X}$ denotes the tuple of variables $x_{hh'}$ for each $(h, h') \in H$.
More explicitly,
$W \vdash  D_{hh'} = \sum_{i = 1}^{k}  a_{i}.D_{f(h,i),f'(h',i)}  + \sum_{j = 1}^{l(h)} \alpha_{hj}.\1$.
If we close each $t_{hh'}$ by replacing $x_{f(h,i),f'(h',i)}$ with $C_{f(h,i)}$, we get

$
\quad \sum_{i = 1}^{k}  a_{i}.C_{f(h,i)} + \sum_{j = 1}^{l(h)} \alpha_{hj}.\1
$

\noindent
which is equal to $body(C_h)$.
Therefore, we note that $C_h$  is 
such that $W \vdash C_h = t_{hh'}\{\widetilde{C}/\widetilde{X}\}$ and
so, by Theorem \ref{unique-w-th}, we have that $W \vdash D_{hh'} = C_h$.
Since $(1, 1) \in H$, we have that $W \vdash D_{11} = C_1$.

Similarly, if we close each $t_{hh'}$ by replacing $x_{f(h,i),f'(h',i)}$ with $C'_{f'(h',i)}$, we get

$
\quad \sum_{i = 1}^{k}  a_{i}.C'_{f'(h',i)} + \sum_{j = 1}^{l(h)} \alpha_{hj}.\1
$

\noindent
which is equal to $body(C'_{h'})$ (as {\bf A1-A2} $\vdash \sum_{j = 1}^{l(h)} \alpha_{hj}.\1 = \sum_{j = 1}^{l'(h')} \alpha'_{h'j}.\1$).  
Therefore, we note that $C'_{h'}$ is 
such that $W \vdash C'_{h'} = t_{hh'}\{\widetilde{C'}/\widetilde{X}\}$ and
so, by Theorem \ref{unique-w-th}, we have 
that $W \vdash D_{hh'} = C'_{h'}$. Since $(1, 1) \in H$, we have that  
$W \vdash D_{11} = C'_1$; by transitivity, it follows that $W \vdash C_1 = C'_1$, and so that $W \vdash p_2 = p_2'$.
Finally, the thesis $W \vdash p = p'$ follows by transitivity.
\fine
\end{lemma}

\begin{theorem}{\bf (Completeness for all the SFM$_0$ processes)}\label{compl-og-ax}

\noindent
For every $p, q \in  \mathcal{P}_{SFM_0}$, if $p \sim q$, then $W \vdash p = q$.

\proof
By Proposition \ref{nf-prop}, there exist $p_1$ and $q_1$
normal forms such that $W \vdash p = p_1$ and $W \vdash q = q_1$. 
By Proposition \ref{satur-prop}, there exist $p_2$ and $q_2$ saturated normal forms
such that $W \vdash p_1 = p_2$ and $W \vdash q_1 = q_2$,
so that $W \vdash p = p_2$ and $W \vdash q = q_2$ by transitivity.

By Theorem \ref{sound-th-sfm0}, we have $p \sim p_1 \sim p_2$ and $q \sim q_1 \sim q_2$,
hence, by transitivity, also $p_2 \sim q_2$.
By Lemma \ref{det-nf-compl}, from $p_2 \sim q_2$ we can derive that $W \vdash p_2 = q_2$, and so, by transitivity, also
$W \vdash p = q$.
\fine
\end{theorem}

%
\section{Conclusion}\label{conc-sec}
%

In this paper, we have shown how to use the process algebra SFM$_0$, that truly represents GFAs up to isomorphism, 
in order to find a sound and complete axiomatization of 
language equivalence, and so also a sound and complete axiomatization of regular languages, 
composed of 7 axioms and 2 conditional axioms, outlined in Table \ref{axiom-sfm0-tab}.

In \cite{Gor23nfa} we have proposed the process algebra SFM1, which truly represent NFAs (with $\epsilon$ transitions), 
up to isomorphism. The axiomatization of language equivalence proposed there contains a variant absorption axiom:

{\bf T3$'$} \;\;($\epsilon$-absorption) \; \; $ \epsilon.x  =  x   \quad  \mbox{ if $x \neq C$}$

\noindent
and one additional conditional axiom:

{\bf R3} \;\;  (Excision) \; \; if $C \eqdef  (\epsilon.x + p)\{C/x\} \wedge D \eqdef p\{D/x\}$ \; then \; $C =  D$.

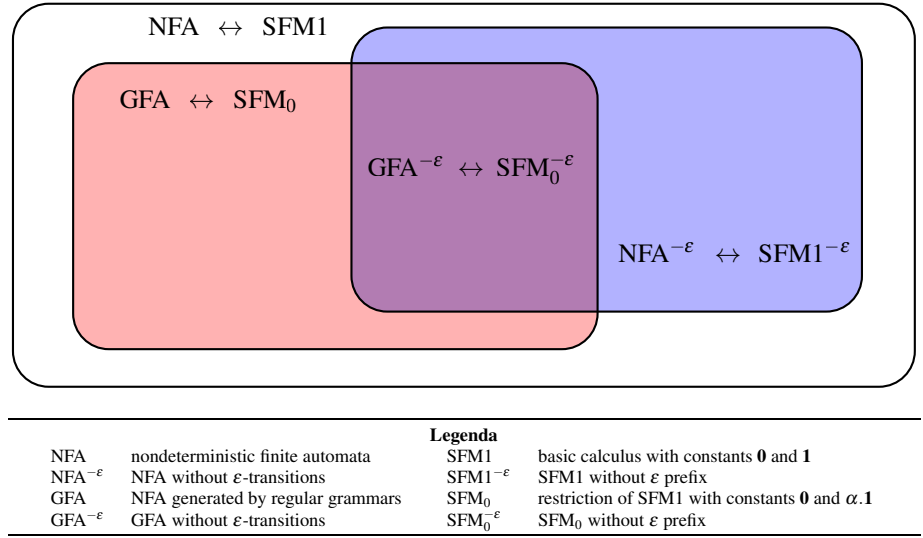
\begin{figure}[ht]
\centering

\begin{tikzpicture}[blend group=multiply]

\draw [rounded corners=3ex, thick] (-2.5,-1.3) rectangle (270pt,24ex);

\draw (0.5,3.5) node {NFA $\; \; \leftrightarrow\; \; $ SFM1};

\draw [rounded corners=3ex, thick,fill=red!30, opacity=0.5] (-1.7,-0.8) rectangle (150pt,19ex);

\draw (0.1,2.5) node {GFA $\; \; \leftrightarrow\; \; $ SFM$_0$};

\draw [rounded corners=3ex, thick,fill=blue!30, opacity=0.5] (2,-0.3) rectangle (250pt,22ex);

\draw (7.1,0.5) node {NFA$^{-\epsilon} \; \; \leftrightarrow\; \; $ SFM1$^{-\epsilon}$};

\draw (3.6,1.6) node {GFA$^{-\epsilon}$ $ \; \leftrightarrow \; $ SFM$_0^{-\epsilon}$};

\end{tikzpicture}

\hrulefill

{\scriptsize
{\bf Legenda}\\
$
\begin{array}{lllllllll}
\mbox{NFA} & \mbox{nondeterministic finite automata} &\quad \mbox{SFM1} & \mbox{basic calculus with constants $\nil$ and $\1$}  \\
\mbox{NFA$^{-\epsilon}$} & \mbox{NFA without $\epsilon$-transitions} & \quad \mbox{SFM1$^{-\epsilon}$} & \mbox{SFM1 without $\epsilon$ prefix}\\
\mbox{GFA} & \mbox{NFA generated by regular grammars} &\quad \mbox{SFM$_0$} & \mbox{restriction of SFM1 with constants $\nil$ 
and $\alpha.\1$}  \\
\mbox{GFA$^{-\epsilon}$} & \mbox{GFA without $\epsilon$-transitions} & \quad \mbox{SFM$_0^{-\epsilon}$} & \mbox{SFM$_0$ without $\epsilon$ prefix}\\

\end{array}
$ 
}

\hrulefill

\caption{Process algebras for finite automata}
\label{riassunto-fig}
\end{figure}

However, if we consider the subcalculus SFM1$^{-\epsilon}$ of SFM1, where $\epsilon$ cannot be used as a prefix, we get that
this can represent all NFAs without $\epsilon$ transitions, a model that can express all the regular languages, too.
The axiomatization of language equivalence over SFM1$^{-\epsilon}$ is a bit 
more compact than that proposed for SFM$_0$
in Table \ref{axiom-sfm0-tab}, as axiom {\bf T3} is not necessary.
Moreover, if we consider the subcalculus SFM$_0^{-\epsilon}$ of SFM$_0$, where $\epsilon$ cannot be used as a prefix, 
we get that it can represent all GFAs without $\epsilon$ transitions, a model that can express all the regular languages that do not contain the empty word. The axiomatization of language equivalence over SFM$_0^{-\epsilon}$ is exactly the same we have proposed above for SFM1$^{-\epsilon}$ (6 axioms and 2 conditional axioms). 
The relations among these process algebras and the classes of finite automata they represent (up to isomorphism) are outlined in Figure \ref{riassunto-fig}.

Other process algebras have been proposed in the literature for modeling NFAs such as those in \cite{Silva,SBR10,BBR09}. A precise comparison between the theory of SFM1
and these is given in \cite{Gor23nfa}.

Our axiomatization of regular languages is alternative
to those based on the Kleene algebra of regular expressions \cite{Salomaa,Conway,Kozen}. As a matter of fact,
our axiomatization, composed of 7 axioms and 2 conditional axioms, seems more concise than these.
In fact, the currently most concise axiomatization of this sort \cite{DDP18} is composed of 13 axioms, 
only one of which is conditional.
However, it should be remarked that our axiomatization makes use of axiom schemata (in particular,
for recursion), while \cite{DDP18} does not.

Another goal of this paper is to show the usefulness
of representability theorems, up to isomorphism (like Theorem \ref{representability-gfa}), for the algebraic 
study of semantic models, like GFAs. In fact, SFM$_0$ has allowed us to study also bisimulation equivalence on GFAs and to 
hint a sound and complete
axiomatization for it (the set $B$), composed of 4 axioms and 2 conditional axioms, only.
Future research may be devoted to extend the axiomatic theory of SFM$_0$ to other behavioral equivalences in 
the linear-time/branching-time
spectrum \cite{vG01,GV15}. For instance, we claim that simulation equivalence \cite{Park81,GV15} can be axiomatized by 
replacing axiom {\bf T2} in the set $W$ with the axiom 

{\bf T2$'$} \quad $a.(x+y) = a.(x+y) + a.y$.

Our axiomatization is based on results and techniques developed in the area of process 
algebras \cite{Mil84,Mil89,HoPA,BBR09,San10,GV15} for bisimulation-based equivalences,
that we have applied and adapted to language equivalence, an equivalence relation
usually not deeply investigated in the context of process algebras.


\section*{References}

\begin{thebibliography}{11}





%
\bibitem{BBR09}
J.C.M. Baeten, T. Basten, M.A. Reniers,
\newblock {\em Process Algebra: Equational Theories of Communicating Processes},
\newblock Cambridge Tracts in Theoretical Computer Science 50, Cambridge University Press, 2010.


%
\bibitem{HoPA}
J.A. Bergstra, A. Ponse, S.A. Smolka (eds.),
\newblock {\em Handbook of Process Algebra},
\newblock Elsevier, 2001.

%
\bibitem{Conway}
J.H. Conway, 
\newblock {\em Regular Algebra and Finite Machines}, 
Chapman and Hall, London, 1971. ISBN 0-412-10620-5. 
 
%
\bibitem{DP02}
B.A. Davey, H.A. Priestley,
\newblock {\em Introduction to Lattices and Order} (second edition),
\newblock Cambridge University Press, 2002.

%
\bibitem{DDP18}
A. Das, A. Doumane, D. Pous, Left-handed completeness for Kleene algebra, via cyclic proofs, 
\newblock in: Gilles Barthe, Geoff Sutcliffe, Margus
Veanes (Eds.), LPAR-22, 22nd International Conference on Logic for Programming, Artificial Intelligence and Reasoning,
EPiC Series in Computing,
vol. 57, pp. 271-289, 2018.

%
\bibitem{vG01}
R.J.~van Glabbeek,
\newblock The linear time - branching time spectrum I,
\newblock Chapter 1 of  \cite{HoPA}, 3-99, 2001.

%
\bibitem{GV15}
R. Gorrieri, C. Versari,
{\em Introduction to Concurrency Theory: Transition Systems and CCS},
EATCS Texts in Theoretical Computer Science, Springer-Verlag, 2015.



%
\bibitem{Gor19}
R.~Gorrieri,
\newblock  Axiomatizing team equivalence for finite-state machines, 
\newblock {\em The Art of Modeling Computational Systems} 
(Catuscia Palamidessi's Festschrift), LNCS 11760, 14-32, Springer, 2019.

%
\bibitem{Gor20ic}
R.~Gorrieri,
\newblock Team equivalences for finite-state machines with silent moves, 
\newblock {\em Information and Computation} Vol. 275, 2020, DOI:10.1016/j.ic.2020.104603

%
\bibitem{Gor23nfa}
R.~Gorrieri,
\newblock The algebra of nondeterministic finite automata, 
\newblock CoRR abs/2301.03435, 2023. https://arxiv.org/abs/2301.03435.



%
\bibitem{HMU01}
J.E. Hopcroft, R. Motwani, J.D. Ullman,
\newblock {\em Introduction to Automata Theory, Languages and Computation}, 2nd ed.,
\newblock Addison-Wesley, 2001.

%
\bibitem{Kleene}
S.C. Kleene,  
\newblock Representation of events in nerve nets and finite automata,
\newblock {\em Automata Studies, Annals of Mathematical Studies}, Princeton Univ. Press. 34 (sect.7.2, p.26-27), 1956.

%
\bibitem{Kozen}
D. Kozen,
\newblock A completeness theorem for Kleene algebras and the algebra of regular events,
\newblock {\em Information and Computation} 110(2):366-390, 1994. doi:10.1006/inco.1994.1037


%
\bibitem{Mil84} R. Milner,
\newblock A complete inference systems for a class of regular behaviors,
\newblock {\em J. Comput. System Sci.}  28: 439-466, 1984.



%
\bibitem{Mil89} R. Milner. {\it Communication and Concurrency},
Prentice-Hall, 1989.



%
\bibitem{Park81}
D.M.R. Park,
\newblock Concurrency and automata on infinite sequences,
\newblock In Proc. 5th GI-Conference on Theoretical Computer Science, LNCS 104, 167-183, 
Springer-Verlag, 1981.

%
\bibitem{RS59}
M.O. Rabin, D. Scott,
\newblock Finite automata and their decision problems, 
\newblock {\em IBM Journal of Research and Development} 3(2):114-125, 1959.


%
\bibitem{Salomaa}
A. Salomaa,
\newblock Two complete axiom systems for the algebra of regular events,
\newblock {\em Journal of the ACM} 13(1): 58-169, 1966. doi:10.1145/321312.321326

%
\bibitem{San10}
D. Sangiorgi, 
\newblock {\em An Introduction to Bisimulation and Coinduction},
\newblock Cambridge University Press, 2012.

\bibitem{Silva}
A. Silva,
\newblock {\em Kleene Coalgebra},
\newblock Ph.D. thesis, Radboud Universiteit Nijmegen, 2010. ISBN: 978-90-6464-433-7

\bibitem{SBR10}
A. Silva, M.M. Bonsangue, and J.J.M.M. Rutten,
\newblock Non-deterministic Kleene coalgebras,
\newblock {\em Logical Methods in Computer Science}, 6(3), 2010.

%
\bibitem{Sud97}
T.A. Sudkamp, 
\newblock {\em Languages and Machines} (second edition),
\newblock Addison Wesley, 1997.

\end{thebibliography}
\end{document}